\documentclass[preprint,12pt]{aastex}
\usepackage{amsmath}

\lefthead{INOUE and CHIBA}
\righthead{Extended Source Effects in Substructure Lensing}
\newcommand{\del}{\partial}

\newcommand\x{{\bf x}}
\newcommand\y{{\bf y}}

\newcommand\bt{\boldsymbol{\theta}}

\newcommand\al{\boldsymbol {\alpha}}

\newcommand\bu{{\bf 1}}
\newcommand\bg{{\bf \Gamma}}

\newcommand{\f}{\frac}
\newcommand{\T}{\tilde}

\newcommand{\bb}{\bibitem}
\newcommand{\BF}{\begin{figure}\begin{center}}
\newcommand{\EF}{\end{center}\end{figure}}
\newcommand{\BE}{\begin{equation}}
\newcommand{\EE}{\end{equation}}
\newcommand{\BEA}{\begin{eqnarray}}
\newcommand{\EEA}{\end{eqnarray}}
\newcommand{\ti}{\textit}

\newcommand{\ms}{M_{\odot}}
\begin{document}

\title{Extended Source Effects in Substructure Lensing}
\author{Kaiki Taro Inoue  \altaffilmark{1} and Masashi Chiba \altaffilmark{2}}
\altaffiltext{1}{Department of Science and Engineering, Kinki University,
Higashi-Osaka 577-8502, Japan;}
\altaffiltext{2}{Astronomical Institute, Tohoku University,
Sendai 980-8578, Japan;}

\email{kinoue@kindai.phys.ac.jp, chiba@astr.tohoku.ac.jp}

%%
%\date{\today}
%%
%\maketitle
\begin{abstract}
We investigate the extended source size effects on gravitational lensing in
which a lens consists of a smooth potential and small mass clumps
(``substructure lensing''). We first consider a lens model that consists of a
clump modeled as a singular isothermal sphere (SIS) and a primary lens modeled
as an external background shear and convergence. For this simple model, we
derive analytic formulae for (de)magnification of circularly symmetric top-hat
sources with three types of parity for their lensed images, namely, positive,
negative, and doubly negative parities. Provided that the source size is
sufficiently larger than the Einstein radius of the SIS, we find that in the
positive (doubly negative) parity case, an extended source is always magnified
(demagnified) in comparison with the unperturbed macrolens system, whereas in
the negative parity case, the (de)magnification effect, which depends on the sign of
convergence minus unity is weaker than those in other parities. It is shown that
a measurement of the distortion pattern in a multiply lensed image enables us to
break the degeneracy between the lensing effects of clump mass and those
 of clump distance if lensing parameters of the relevant macrolens model are
determined from the position and flux of multiple images. We also show that an
actual density profile of a clump can be directly measured by analyzing
 the ``fine structure'' in a multiply lensed image within the Einstein
 radius of the clump.

\end{abstract}

\section{Introduction}
The standard cold dark matter (CDM) scenario predicts 
the presence of several hundred small-mass 
clumps or ``subhalos'' ($\lesssim 10^8 \ms$)
in a galaxy-sized halo ($\sim 10^{12} \ms$), while the observed
number of dwarf galaxies around the Milky Way is only a dozen
(Klypin et al. 1999; Moore et al. 1999). This suggests the presence of several hundred subhalos holding few or no stars around such a galaxy.
Gravitationally lensed QSOs with quadruple images have recently
been used for putting a limit on the surface density and the mass of such
invisible substructures (Mao \& Schneider 1998; Metcalf \& Madau
2001; Chiba 2002; Metcalf \& Zhao 2002; Dalal \& Kochanek 2002; 
Bradac et al. 2002). Statistical analyses on these quadruple QSO-galaxy 
lensing systems show that the (de)magnification of lensed 
images and its dependence on their
parities can be explained by the presence of substructures along
the line of sight to the images, in contrast to the lens models relying on
``smooth perturbation'' in the gravitational 
potential of a macrolens
(Metcalf \& Zhao 2002; Keeton et al. 2002; Evans \& Witt 2003).

However, there are two ambiguities in these QSO-galaxy lensing systems.
First, the subhalo masses are poorly measured. We can put 
meaningful constraints on these masses by observing 
substructure lensing of extended sources with various sizes (Moustakas \& Metcalf 2003; 
Dobler \& Keeton 2005). 
If the angular source size is sufficiently 
larger than the Einstein radius of the subhalo, then the images lensed
by a smooth galaxy halo will not be perturbed by 
a subhalo at all. 
Thus, by using a source with a large 
angular size, we can put a stringent constraint 
on the lower limit of the subhalo mass. 
Various examples of an extended
source are available in QSO-galaxy lensing systems.
For instance, a QSO core component has a typical size of
$\lesssim 1 \times (\nu/5 \textrm{GHz})^{-1}$~pc in the radio band
(Kameno et al. 2000;
Kadler et al., 2003), which is to be compared with the Einstein radius of
$\sim 1 \times (M/(3\times 10^5 M_\odot))^{1/2}$~pc for 
a point-mass clump with mass $M$.
In addition, a surrounding hot dust torus around a QSO nucleus is supposed to be
as large as $0.1-1$~pc, and the blackbody emission from this hot dust is
observable in the infrared band (e.g., Agol, Jones, \& Blaes 2000;
Chiba et al. 2005).
Cold dust components at temperatures of $20-50$K around a QSO nucleus can be as
large as several kpc (e.g., Puget et al. 1996).

Second, the locations of subhalos along the line of sight are poorly
constrained. In fact, extragalactic substructures other than 
those associated with primary lensing
halos can also contribute significantly to QSO-galaxy lens systems.
Using $N$-body simulations, Wambsganss et al. (2004) showed that
the fraction of cases for which more than one lens plane contributes
significantly to the multiply imaged system is very large for a source
at high redshift $z\gtrsim 3$. Metcalf (2005) showed that all the cusp caustic
lens anomalies can also be explained by extragalactic $\Lambda$CDM halos with
a mass range of $10^{8}-10^{9}\ms$. 
In order to determine the distances to these substructures, we need to have
more information in addition to the position and flux of multiple images.
Future high-resolution 
mapping of extended images in a QSO-galaxy lensing 
system may provide us useful information 
about the masses, distances,
and density profiles of such substructures (Inoue \& Chiba 2003, 
2005). Such an observation can be achieved by 
next-generation space VLBI such as the 
VLBI Space Observatory Program 2 (VSOP2; Hirabayashi et al. 2001) and submillimeter
interferometers such as the Atacama Large Submillimeter Array (ALMA). 

To address these issues, we need to understand 
substructure lensing effects for an extended 
source. However, to date little attention 
has been paid to the effects of a source size that is
comparable to or larger than the Einstein 
radius of a perturber although lensing of 
such an extended source by a point mass 
(i.e., without
a background, primary lens) has been studied extensively 
(e.g., Witt \& Mao 1994). 
In comparison with lensing systems in which the primary lens is a cluster of
galaxies, a galaxy-sized halo can be represented by a relatively small number
of parameters since it has many symmetries. This allows us to treat
the lensing effect of small-mass substructures as a local 
perturbation to
the macrolensing effect in QSO-galaxy lensing systems.
For point sources, the dependence on image parities of their (de)magnification
has been investigated (Finch et al. 2002; Keeton 2003).
It is of great importance to know to what extent the finite source size
affects such systematic (de)magnification in substructure lensing.

In this paper, we investigate the extended source effects in substructure
lensing. 
In section 2 and 3, we show that 
for an SIS lens (with or without background shear and convergence), 
the flux of a circular source can be significantly altered 
even if the source size is much larger than that of the 
Einstein ring. In section 2, we consider
a single SIS model without background for 
a circularly symmetric top-hat or Gaussian
surface brightness profile of a source.
In section 3, we explore SIS lens systems with 
an external background shear $\gamma$ and convergence $\kappa$ 
relevant to substructure lensing.
We derive analytic (de)magnification formulae based on 
astrometric shifts 
for a circular symmetric top-hat source 
for three types of parity of an image, namely, positive,
negative and doubly negative parities.  We show that 
in the large source size limit, 
the (de)magnification effect depends systematically on the parity of an 
the image, and it is prominent even for 
sources somewhat larger than the Einstein ring. 
We also discuss the mean 
(de)magnification perturbation for a point source.   
In section 4, we study a simple
method of breaking the degeneracy 
between the lensing effects of subhalo mass and distance 
using astrometric shifts of a macrolensed image. 
In section 5, we explore 
a method to directly measure the mass density 
profile of each lens perturber. 
In section 6, we summarize our results.

\section{SIS lens without background}
In this section, we show that 
for an SIS, the magnification effect is still prominent
even if the source size is larger 
than the Einstein radius. This implies that if 
the source size dependence 
of magnification is measured, 
then one can determine the Einstein radius of an SIS
lens if the source center is fixed. Note that
an SIS is widely used in the literature for representing
the density profile of 
CDM subhalos with $10^{7}-10^{9}\ms$ relevant to
substructure lensing (Metcalf \& Madau 2001; Dalal \& Kochanek 2002), 
although it is somewhat more concentrated than the so-called 
Navarro-Frenk-White (NFW) profile
(Navarro, et al. 1996) near the central cusp.
For simplicity, we ignore the effect of a background shear and
convergence as well as the ellipticity of the lens, in this section.

Let $\bt_x=(\theta_{x1},\theta_{x2})$ and 
$\bt_y=(\theta_{y1},\theta_{y2})$ 
be two-component vectors 
of angles in the sky representing the image position and
the source position, respectively. 
The lens equation for an SIS is then
\BE
\bt_y=\bt_x- \theta_E \f{\bt_x}{|\bt_x|},
\EE
where the Einstein radius $\theta_E$ is written in terms of
velocity dispersion $\sigma $ and light velocity $c$ as 
\BE
\theta_E=4 \pi \biggr( \f{\sigma}{c} \biggl)^2 \f{D_{LS}}{D_S}, 
\EE
where $D_{LS}$ and $D_S$ are angular diameter distances
between the lens and the source, and between the observer and the source,
respectively. In what follows, we set $\theta_E=1$ without loss
of generality.
 In terms of complex variables 
$z=\theta_{x1}+i \theta_{x2}$ and  
$\zeta=\theta_{y1}+ i \theta_{y2}$, the normalized lens equation
is expressed as  
\BE
\zeta=z-\f{z}{|z|},
\EE
which has a set of solutions
\BE
z_{+}=(1+|\zeta|)\f{\zeta}{|\zeta|}
,~\textrm{for}~~\forall~ \zeta,~~
z_{-}=(|\zeta|-1)\f{\zeta}{|\zeta|}
,~\textrm{for}~~|\zeta|<1.
\label{eq:z}
\EE
corresponding to an image position with 
positive parity (denoted by the subscript plus sign) 
and that with negative parity (denoted by the subscript minus sign).

\subsection{Top-hat source}

In this subsection, we explore a 
circularly symmetric top-hat source 
with a radius $L$ in units of the Einstein radius
with a constant surface brightness,
located at a distance $\zeta_0$ from the center of an SIS lens.

In terms of a complete elliptic integral of the first kind $K(k)$ and
that of the second kind $E(k)$, the 
magnification factors of a circular top-hat source 
corresponding to either 
a positive parity image $z_+$ or a negative parity 
image $z_-$ can be written as 
\BE
\mu_{\pm}(L,\zeta_0)=\Biggl| 1\pm \f{2}{\pi L^2}
\biggl((L+\zeta_0)E(k)+(L-\zeta_0)K(k)  
\biggr)\Biggr|,~~ k=\Biggl(\f{4 L \zeta_0}{(L+\zeta_0)^2}
\Biggr)^{1/2}. \label{eq:mu3}
\EE

To understand the effect of the finite source radius $L$, first,
we study the case in which a circular top-hat source is centered
at the lens center ($\zeta_0=0$).
From equation (\ref{eq:mu3}), one can find that
the magnification factor $\mu=\mu_{-}$
diverges as $\mu \sim 4/L $
in the small size limit $L\rightarrow 0$, similar to that of a point 
mass lens with a unit Einstein radius, 
$\mu \sim 2/L$ (Witt \& Mao 1994). 
However, in the large-size limit $L\rightarrow \infty$, 
the magnification factor
for an SIS lens converges more slowly to unity,   
as $\mu\sim \mu_+ \sim 1+2/L+1/L^2$, than that of a point mass lens,  
$\mu\sim 1+2/L^2$ (Witt \& Mao 1994).
Therefore, for an SIS, the magnification effect cannot be negligible 
even if the source size is greater 
than the Einstein radius. 
For instance, the source flux can be 
altered by 20 percent even if the source size is 10 times 
larger than the Einstein radius.
This is because the deviation of the photon path coming from regions
outside the Einstein radius is larger for an SIS lens
than that for a point mass lens with the same
Einstein radius. This behavior is reasonable,
since the deflection angle for an SIS lens is
constant with increasing $|\bt_x|$, whereas
it decays as $1 / \bt_x$ for a point mass lens.
%since the two-dimensional 
%gravitational potential in the lens plane decays 
%slowly as $\Phi(x)\propto 
%\textrm{ln}(x)$ for an SIS lens
%whereas it decays as $\Phi(x)\propto 
%-1/x$ for a point mass lens.

Next, we study the case  
in which a circular top-hat source is placed at 
an off-center position $(0,\zeta_0 \ne 0)$ in the lens plane. 
For $L>1 and \zeta_0 > 1$,
the magnification factor can be approximately 
given by (see Appendix A)
\BE
\mu(L,\zeta_0)\approx \mu_{+}(L,\zeta_0)+\delta \mu(s,L,\zeta_0),~~~ 
L>1,\zeta_0>1. \label{eq:appmu}
\EE
where
\BE
\delta \mu(s,L,\zeta_0)=\f{\textrm{Erf}(s (L-\zeta_0))
- \textrm{Erf}(-s (L-\zeta_0))+2}{4 L^2}, \label{eq:appf}
\EE
where $s>0$ controls smoothness of $\delta \mu$ around $L=\zeta_0$.
As shown in Figure 1, the magnification 
factor $\mu$ keeps its value 
$\mu\sim 1+1/\zeta_0$ 
until the boundary of the source touches 
the edge of an Einstein ring. Then
it starts to increase until the source completely
includes the Einstein ring adding an extra 
contribution $\delta \mu=1/L^2$ corresponding to
the ``second'' image, which is shown in 
the top right and bottom left panels in Figure 2. 
As the source radius $L$ increases, 
$\mu$ gradually decreases as $\mu \sim \mu_++1/L^2$. 
Because $\mu_+ \sim 1+2/L$ for $L \gg1 $,  the magnification factor
can again be significantly altered even if the source size is 
larger than that of the Einstein ring provided that the 
Einstein ring is totally ``occulted'' by the source. 
\begin{figure*}
\epsscale{0.7}
\plotone{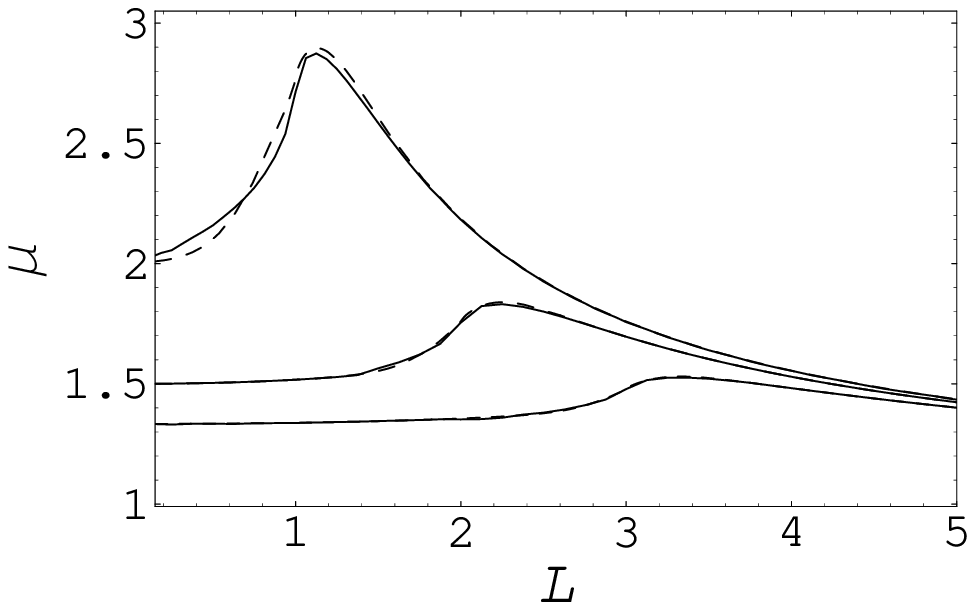}
\label{figmu}
\caption{Magnification factor $\mu$ ({\it{solid curves}}) for 
a top-hat circular source lensed by an SIS  
without background 
as a function of a source radius $L$ with distance from lens center 
$\zeta_0=1$ ({\it{top curves}}),
$\zeta_0=2$ ({\it{middle curves}}) and $\zeta_0=3$ 
({\it{bottom curves}}) in comparison
with analytically calculated values based on eq.
(\ref{eq:appmu}) with smoothing parameter $s=3$ ({\it{dashed curves}}). }
\end{figure*}

\begin{figure*}
\epsscale{1}
\plotone{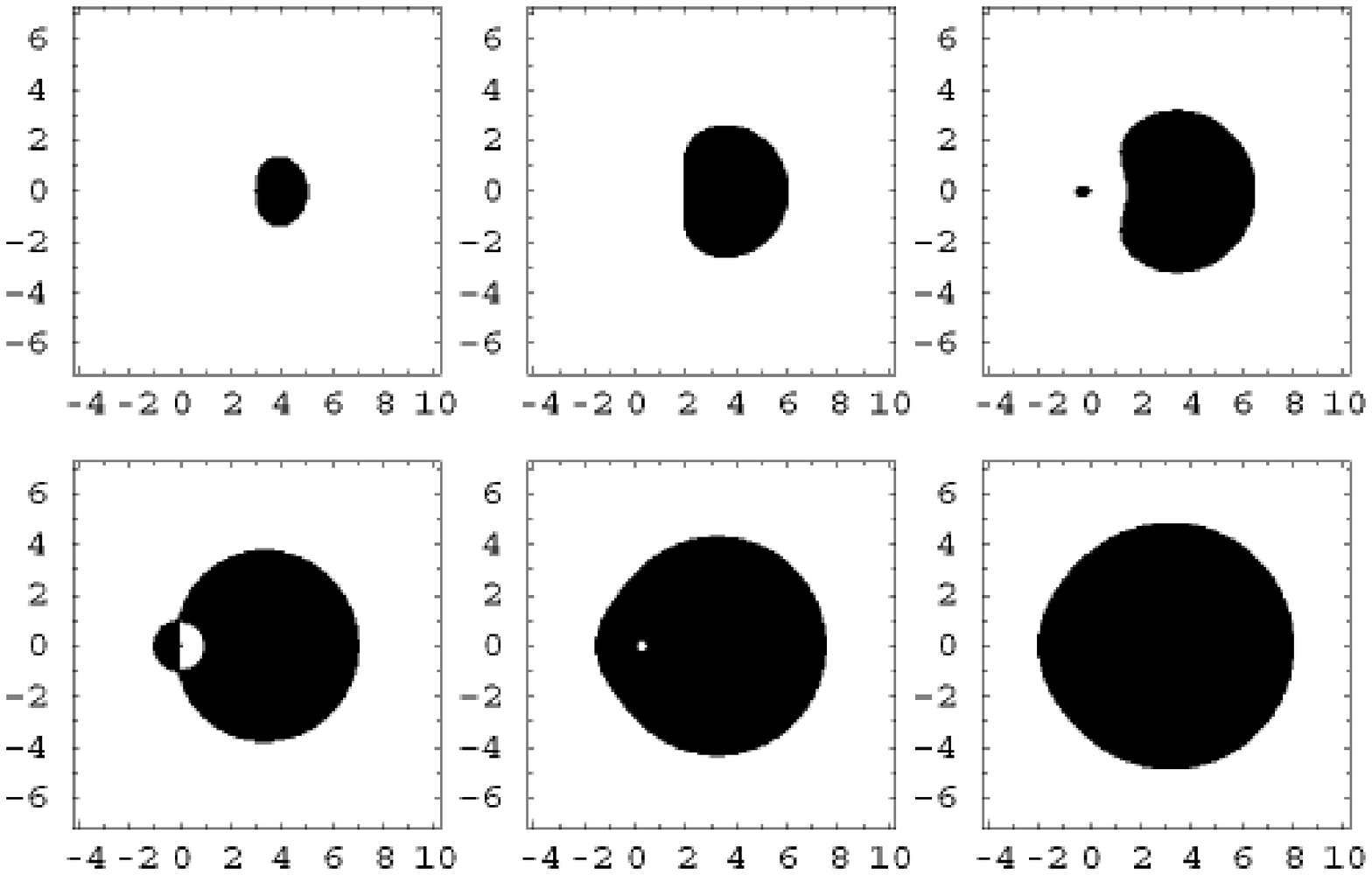}
\label{image}
\caption{Lensed images of a top-hat
circular source centered at $\zeta=(3,0)$ 
with a radius $L=1$ ({\it{top left}}), $L=2$ ({\it{top middle}}),
$L=2.5$ ({\it{top right}}), 
$L=3$ ({\it{bottom left}}), $L=3.5$ ({\it{bottom middle}}), and
$L=4$ ({\it{bottom right}}) 
in the source plane. An SIS lens is centered at
$(0,0)$ in the lens plane. The Einstein radius is set to unity.}
\end{figure*}

\subsection{Gaussian source}
In this subsection, we explore a circularly symmetric Gaussian source 
located at distance $\zeta_0$ from the center of an SIS lens.
The surface brightness of a Gaussian circular source 
with standard deviation $L$ is
\BE
f(L,\zeta)=\f{1}{2 \pi L^2} \exp \biggl [- \f{|\zeta|^2}{2 L^2} 
\biggr ].
\EE
The magnification factor for a Gaussian circular source 
is then
\BE
\mu^G= \biggl|\int_{\textrm{all}} f(L,\zeta(z)) d z
\biggr|. \label{eq:muG}
\EE
\begin{figure*}
\epsscale{0.7}
\plotone{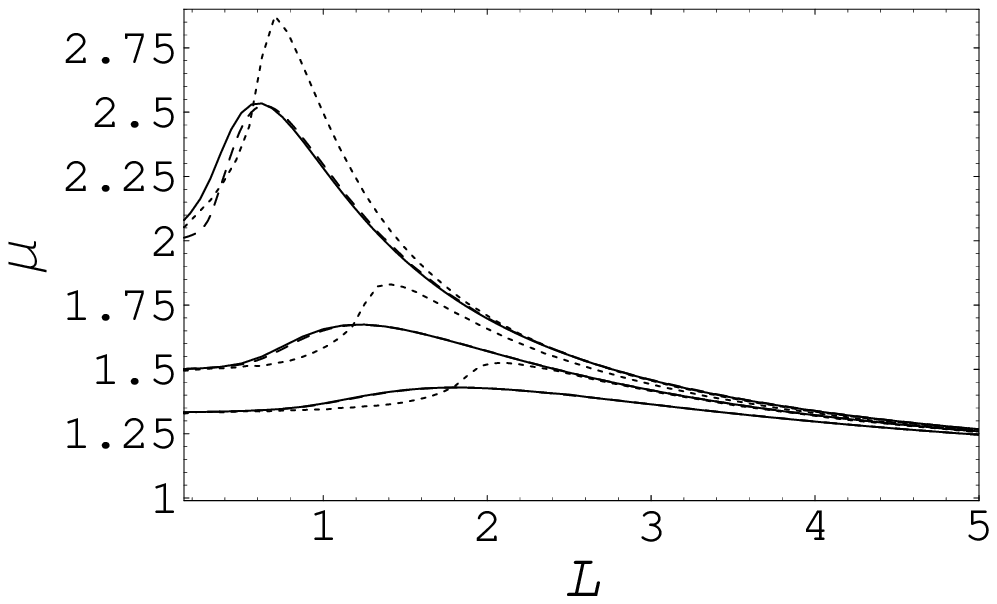}
\label{figmuG}
\caption{Magnification factor $\mu$ ({\it{solid curves}}) for 
a Gaussian circular source lensed by an SIS without background 
as a function of a standard deviation $L$ with distance 
from the lens center $\zeta_0=1$ ({\it{top curves}}),
$\zeta_0=2$ ({\it{middle curves}}), and $\zeta_0=3$ 
({\it{bottom curves}}) in comparison
with those for an image with positive parity only 
(see eq. (\ref{eq:mugp}); {\it{dashed curves}}) and
those for top-hat sources with a scaling 
$L \rightarrow \sqrt{8/\pi} L$ (here $L$ denotes a 
radius of a circular source) for the same $\zeta_0$ 
values ({\it{dotted curves}}).}
\end{figure*}
\begin{figure*}
\epsscale{0.9}
\plotone{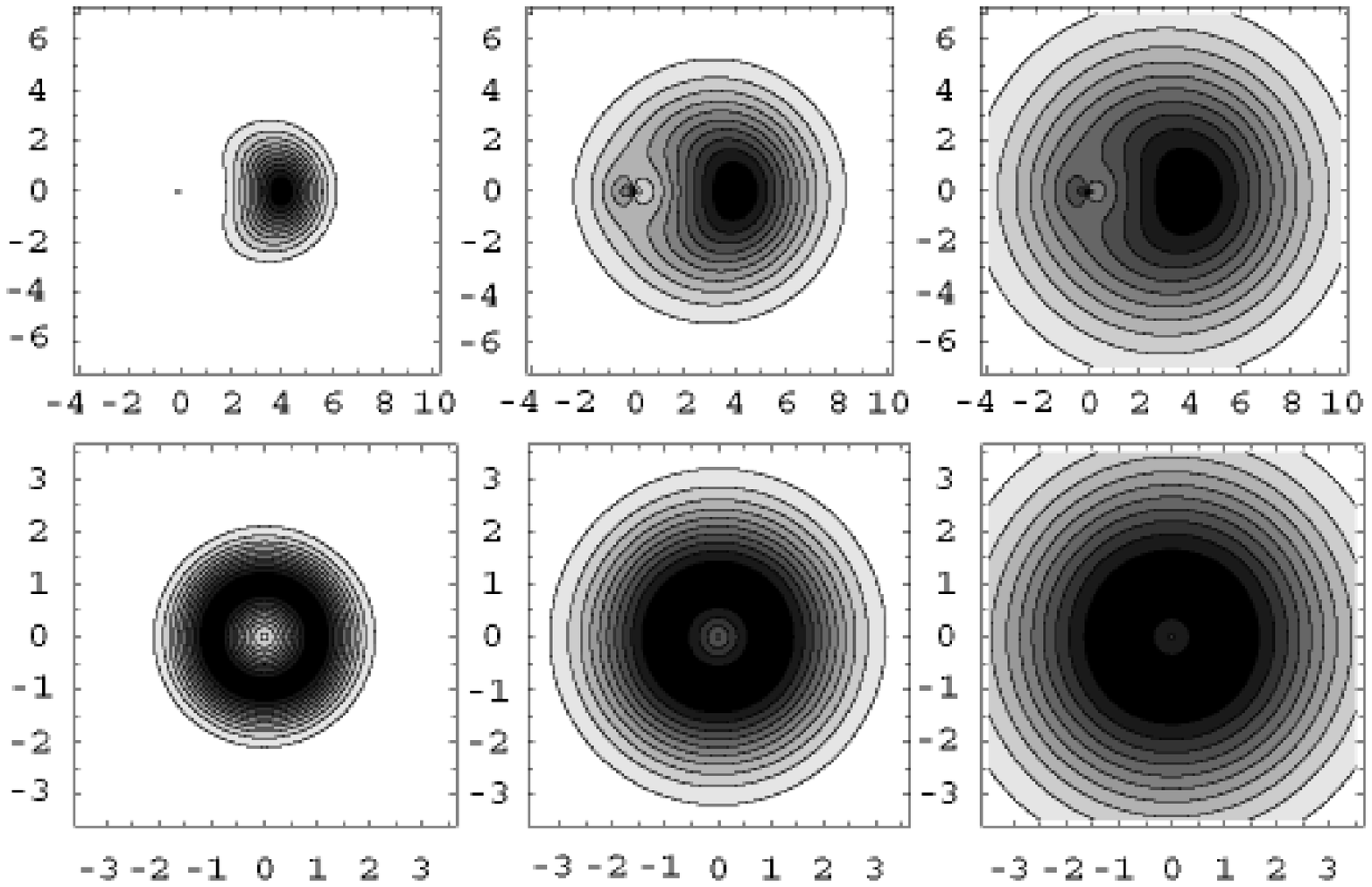}
\label{image-gauss}
\caption{{\it{Top}}: Lensed images 
of a Gaussian circular source centered at $\zeta=(3,0)$
with a standard deviation $L=1$ ({\it{left}}), $L=2$ ({\it{middle}}), 
and $L=3$ ({\it{right}}). {\it{Bottom}}: Lensed images
with the same parameters except for the position of 
the source at $\zeta=(0,0)$. An SIS lens 
is centered at $(0,0)$.
Each interval between 
neighboring contours is 1/10 of the maximum value
in the surface brightness. The Einstein radius is set to unity. }
\end{figure*}
For an image with positive parity, equation (\ref{eq:muG}) can be
reduced to 
\BE
\mu_+^G=-\f{1}{2 \pi L^4}  \int_{0}^\infty d R \int_{0}^{2 \pi}
 d \phi ~  \textrm{Re}~ z_+ \f{\del \textrm{Im}~z_+}{\del \phi} R \exp \biggl[-\f{R^2}{2 L^2}
  \biggr].   
\label{eq:mugp}
\EE
Expanding equation (\ref{eq:mugp}) in $\zeta_0$, we 
obtain the magnification factor $\mu$ in the large source size 
limit, $L \rightarrow \infty$, as
\BE 
\mu^G\approx 1+(\pi/2)^{1/2}L^{-1}+L^{-2}/2.
\EE
As one can see in Figure 3, the 
contribution from an image with negative parity is negligible
in the large source size limit. As in the top-hat model, the
order of the leading term in $\mu-1$ is ${\cal{O}} (L^{-1})$.
Therefore, the magnification effect for a Gaussian source cannot be negligible 
even if the source size is greater 
than the Einstein ring.

To estimate $\mu $ in the small source size 
limit $L \ll 1$, we introduce
a cutoff radius $R_{\textrm{cut}}\sim L$ such that 
the contribution of integrand in equation (\ref{eq:mugp}) 
for $R>R_{\textrm{cut}}$ is negligible. 
If we further assume that $R_{\textrm{cut}} \ll 1<\zeta_0$, then
equation (\ref{eq:mugp}) yields
\BE
\mu^G \sim \f{R_{\textrm{cut}}^4}{2 L^4} \biggl( 1+\f{1}{\zeta_0}
 \biggr ).
\EE 
In the limit $\zeta_0 \rightarrow \infty$, no magnification is 
expected, i.e., $\mu=1$.
Therefore, we obtain the same formula for the small source size
limit, $\mu^G \sim 1+ \zeta_0^{-1}$ for $L\ll 1$ and $\zeta_0 
>1$, as in the top-hat model.
  
As one can see in Figure 3, for $L\gg\zeta_0+1$
the behavior of the 
magnification factor for a Gaussian source 
$\mu^G$ as a function of $L$ is 
similar to that of $\mu$ for a top-hat source, 
if an appropriate scaling for the source radius is carried out.
Therefore, for an SIS, we conclude that the surface brightness 
profile does not play an important role in 
magnification in the large source size limit,  
provided that the source has a circular symmetry. 

However, in the Gaussian model, distorted images
have two distinct features depending on the position of the SIS
: (1) a pair of bright and dark spots
(dipole structure) at the position of the SIS where the spatial
gradient in the surface brightness $\nabla f$ does not 
vanish (Figure 4, {\it{top}}), and (2) a dark spot (monopole structure) at the position of the SIS
at the peak in the surface brightness where $\nabla f=0$ 
(Figure 4, {\it{bottom}}). 
As we see in section 4 and 5, these 
structures are not specific to SIS models. By looking into
these small-scale ``fine structures'', one can measure
the density profile and the mass of the lens perturber. 

\section{SIS lens $+$ background shear and convergence}
In this section, we show that in the large 
source size limit, the order of the (de)magnification perturbation 
of a circular top-hat source with radius $L$ (in units of an 
Einstein radius) owing to an SIS subhalo is 
typically ${\cal{O}}(L^{-1})$ with respect to that for an 
unperturbed lens. We model lensing by a subhalo locally as 
a background constant shear and 
convergence (Finch et al. 2002; Keeton 2003). 
If the mass scale of 
a substructure is sufficiently smaller than that of the macrolens,
this model provides a good approximation in representing the 
local property of lensing along the line of sight to a substructure.
For simplicity, we study a circular top-hat source 
centered at the SIS lens center.   

\subsection{Systematic distortion} 
First, we study the systematic distortion of a 
circular source.  
Let us consider an SIS lens with a one-dimensional 
velocity dispersion $\sigma$ and an external 
convergence $\kappa$ and shear $\gamma>0$. In the coordinates
aligned to the shear, the lens equation is 
\BE
\bt_y=(\bu-\bg)\bt_x-\theta_E \f{\bt_x}{|\bt_x|},
\EE
where
\BE
\bg= \left[\begin{array}{cc}
    \kappa+\gamma &   0   \\
      0   & \kappa-\gamma \\
    \end{array}\right] 
%\begin{pmatrix}
%\kappa+\gamma  & 0 \\ 0&\kappa-\gamma
%\end{pmatrix}
\EE
and
\BE
\theta_E=4\pi \f{\sigma^2}{c^2} \f{D_{LS}}{D_S},
\EE
where $D_S$ is the angular diameter distance to the
SIS lens, $D_{LS}$ is the distance between the SIS and the source,
and $c$ is the light velocity.
In terms of normalized coordinates, $\x=\bt_x/\theta_E$
and $\y=\bt_y/\theta_E$, the
lens equation can be reduced to 
\BE
\y=(\bu-\bg)\x-\f{\x}{|\x|}. \label{eq:lens}
\EE
To characterize the effect of background shear and 
convergence, we examine how a set of orthogonal vectors
$(y_1,0),(0,y_2)$ in the source plane is mapped 
to another set of orthogonal vectors
$(x_1,0),(0,x_2)$ in the lens plane that is aligned to the
shear (\ti{i.e.}, astrometric shifts). 
From the lens equation 
(\ref{eq:lens}), we have 
\BEA
x_1 &=&\biggl (y_1+\f{x_1}{|x_1|} \biggr)
(1-\kappa-\gamma)^{-1}
\nonumber
\\
x_2 &=&\biggl (y_2+\f{x_2}{|x_2|} \biggr)
(1-\kappa+\gamma)^{-1}. \label{eq:axes}
\EEA
On the one hand, 
because of equation (\ref{eq:axes}), the presence of an SIS perturber
leads to a shift of a horizontal vector of the 
unperturbed image $(x_1>1,0)$ to $(x_1+1/(1-\kappa-\gamma),0)$, 
and a shift of a vertical vector of the 
unperturbed image $(0,x_2>1)$ to $(0,x_2+1/(1-\kappa+\gamma))$. 
On the other hand, as we have seen in section 2, 
for $\kappa=\gamma=0$, a circular top-hat 
source is isotropically magnified to a 
circular top-hat image with radius $1+L$.
From these properties, we can expect that  
the boundary of the perturbed macrolensed image 
takes the form of an elliptic with 
a semi-major axis $a+1/(1-\kappa-\gamma)$ 
and a semi-minor axis $b+1/(1-\kappa-\gamma)$.
The validity of this approximation 
is discussed in section 3.2.

Based on this approximation,  
distortion patterns are classified into three regimes by 
the sign of the eigenvalues of $\bu-\bg$,
$\xi_1\equiv1-(\kappa+\gamma)$ and 
$\xi_2\equiv1-(\kappa-\gamma)$:$ (1)$ 
$\xi_1>0, \xi_2>0$ positive parity;
(2) $\xi_1<0, \xi_2>0$
negative parity; and (3)
$\xi_1<0, \xi_2<0$ doubly negative parity\footnote{
These parities correspond to a minimum (positive), a saddle (negative), 
 and a maximum (doubly negative) point in the arrival time surface. }. 
As shown in Figure 5, in 
the positive parity case, both the semi-major axis 
and the semi-minor axis of an image extend
by $\Delta a=\xi_1^{-1}$
and $\Delta b=\xi_2^{-1}$ in comparison with
the unperturbed lensed image (background shear and convergence only)
while in the doubly negative parity case,  
both axes shrink, i.e., $\Delta a<0$ and $\Delta b<0$.
In the negative parity case, the semi-major axis shrinks
by $|\Delta a<0|=|\xi_1^{-1}|$ whereas the semi-minor axis 
extends by $\Delta b=\xi_2^{-1}$ (Figure 6).
Note, however, that a deviation from 
an elliptic is no longer negligible in the negative parity case, 
which is discussed in section 3.2.

\begin{figure*}
\epsscale{1}
\plotone{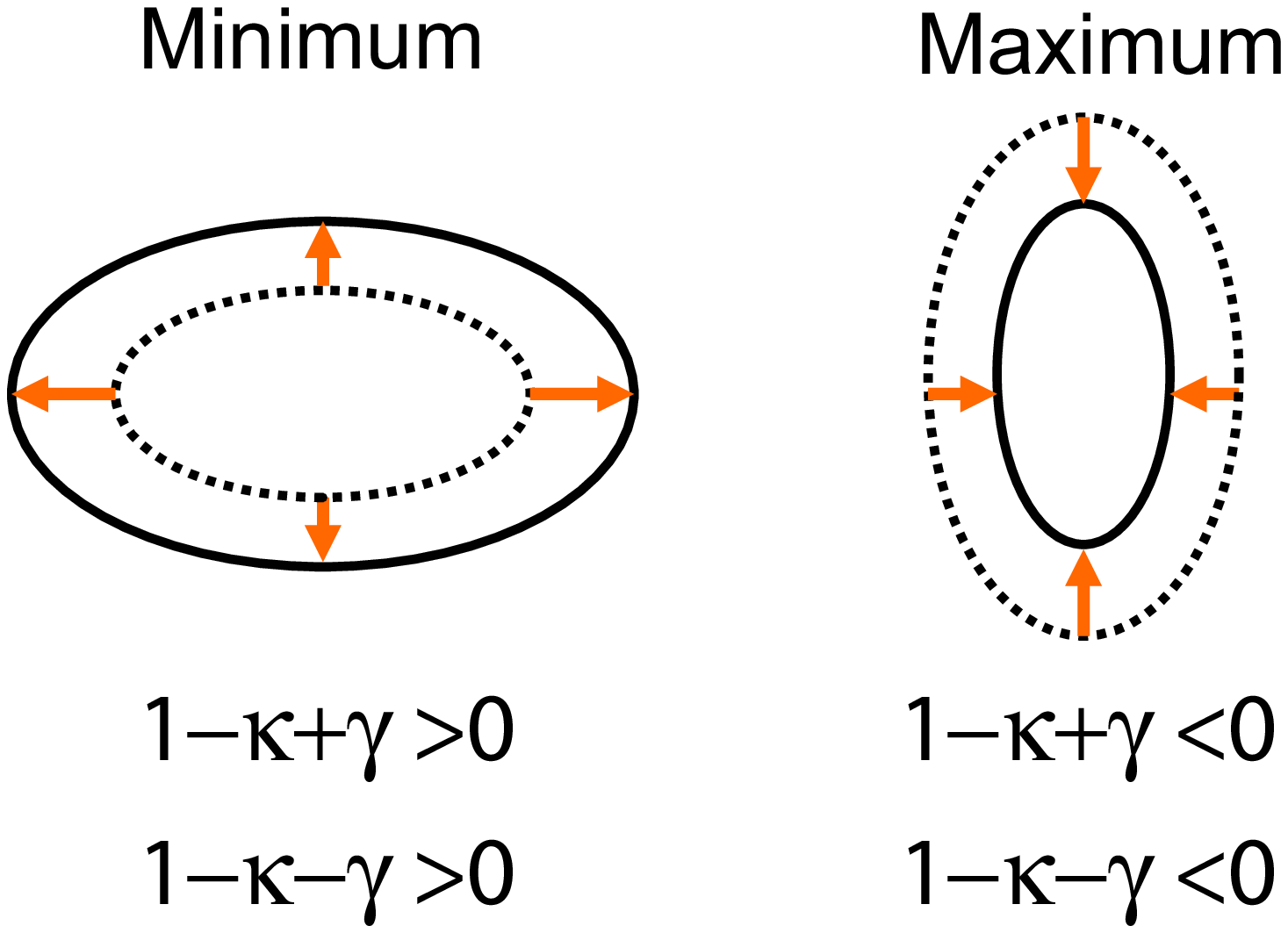}
\caption{Systematic distortion by an 
SIS perturber at the center of a top-hat circular source
with radius $L \gg 1$ in an external shear $\gamma$ and 
convergence $\kappa$.
The dotted and solid closed curves 
show the boundaries of an unperturbed and a perturbed 
macrolensed image, respectively.  
}
\label{fig:distortionNa}
\end{figure*}

\begin{figure*}
\epsscale{1}
\plotone{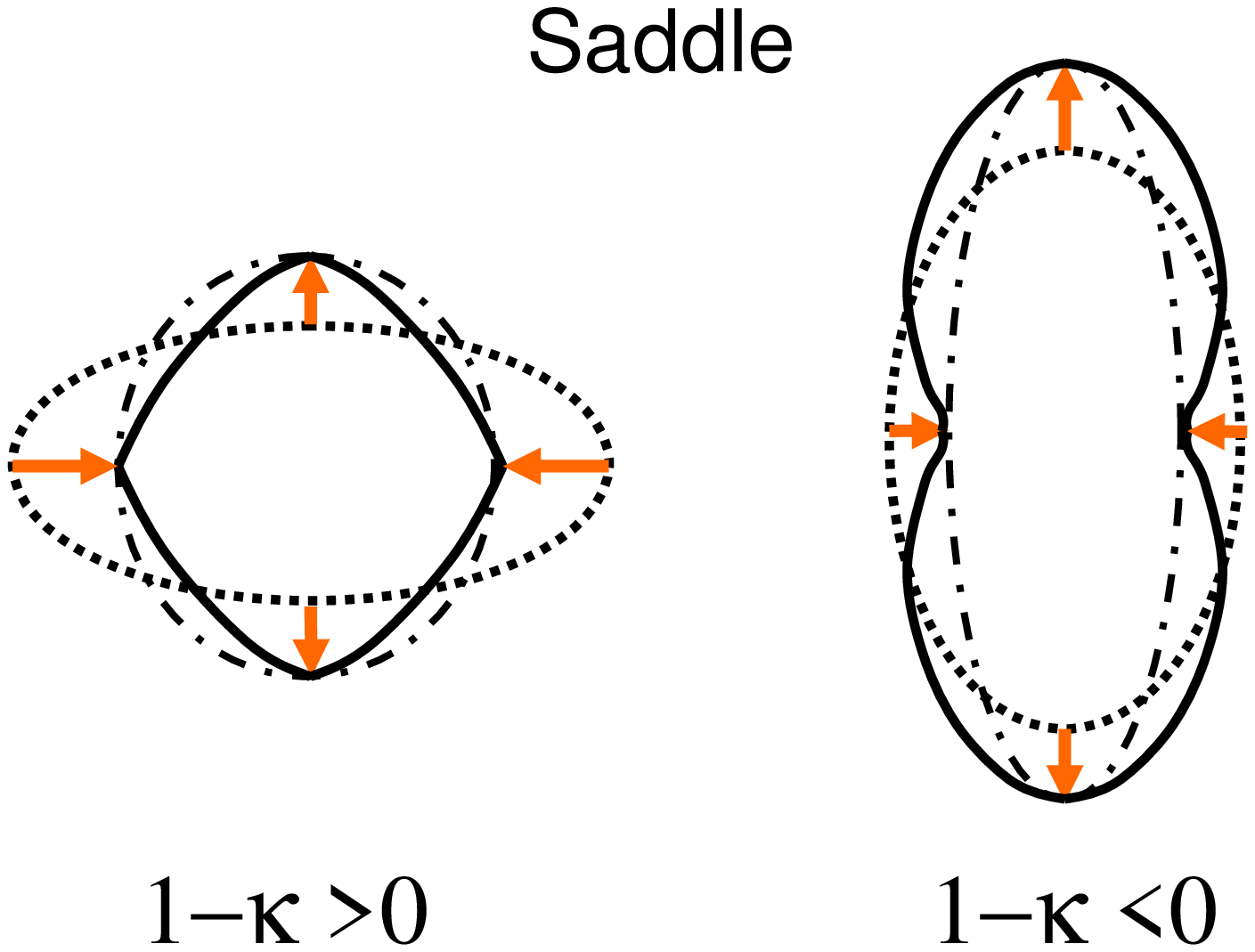}
\caption{Systematic distortion by an SIS perturber at the center of a top-hat circular source 
in an external shear $\gamma$ and 
convergence $\kappa$.
The dotted and solid closed curves 
show the boundaries of an unperturbed and a perturbed 
macrolensed image, respectively.
The dash dotted closed curves are boundaries of an ellipse
with a semi-major axis $a=(L-1)\xi_1^{-1}$ and
a semi-minor axis $b=(L+1)\xi_2^{-1}$. The total area
inside the boundary gets smaller (larger) than the unperturbed 
macrolensed image for $1-\kappa>0$ (for $1-\kappa<0$)
provided that $L \gg 1$.
}
\label{fig:distortionNb}
\end{figure*}

\subsection{Systematic (de)magnification} 
In this subsection, we examine systematic (de)magnification 
for a lens that consists of an SIS plus a constant shear and a 
convergence that depends on the sign of the eigenvalues of
$\bu-\bg$. For simplicity, first, we consider
a circularly symmetric top-hat source with radius $L\gg 1 $
centered at the position of an SIS. Next, we compare our
analytically calculated (de)magnification values 
with numerically calculated ones in cases where the source is
placed at an off-center position. 

First, we study the magnification effects in the positive parity case
and the doubly negative parity case.
As shown in Appendices B and C, in these cases, 
the shape of a distorted image perturbed by an SIS is 
well approximated by an ellipse, provided that $L\gg1$ and 
$|1-\kappa|-\gamma \gg 0 $. Therefore,
a (de)magnification ratio defined as 
the ratio of the magnification for a perturbed lens 
(perturber + macrolens) to the magnification for an unperturbed caustics
lens (macrolens only) $\chi(L)=\mu/\mu_{\textbf{bg}}$
can be simply written as
\BE
\chi(L)\approx 1+\f{2 \textrm{sgn}(\xi_1)}{L}+\f{1}{L^2}. 
\label{eq:pp-nn}
\EE
Thus, the order of the leading term in the 
(de)magnification perturbation defined as 
$\chi-1$ is $\sim {\cal{O}}(L^{-1})$.   
A perturbed image is always magnified (demagnified)
in comparison with the unperturbed image in the 
positive (doubly negative) parity case.
As shown in Figure 7 ({\it{left}}), the 
analytic formula (\ref{eq:pp-nn}) 
for a circular source centered at an SIS is accurate to 
within a few percent for $L\gg 1$. 
Even if a circular source 
is not centered at the SIS 
perturber, formula (\ref{eq:pp-nn}) still gives 
good accuracy for $L\gg 1$ as shown in Figure 8.

Next we consider the negative parity case. If the shape of an
image perturbed by an SIS is well approximated by an 
ellipse, the (de)magnification ratio should be $\chi(L)=1-1/L^2$. However, in the negative
parity case, distortion from an elliptic shape is no longer negligible
at ${\cal{O}}(L^{-1})$.
To quantify departure from an ellipse, we introduce
two parameters $\epsilon(\theta)$ and $\eta(\theta)$ 
in the lens equation in polar coordinates that satisfy
\BEA
L \cos(\theta+\eta)&=&a(1+\epsilon)(1-\kappa-\gamma)\cos \theta
-a \cos \theta /R[a,b,\theta]
\nonumber
\\
L \sin(\theta+\eta)&=&b(1+\epsilon)(1-\kappa+\gamma)\sin \theta
- b \sin \theta/R[a,b,\theta], \label{eq:lens-approx2}
\EEA
where
\BE
a=(L+\textrm{sgn}(\xi_1))\xi_1^{-1}, 
b=(L+\textrm{sgn}(\xi_2))\xi_2^{-1}, 
\EE
and
\BE
R[a,b,\theta]=(a^2 \cos^2 \theta +b^2 \sin^2 \theta)^{1/2}.
\EE

In terms of $\epsilon$, the ratio of the 
perturbed magnification factor to the 
unperturbed magnification factor can be approximately 
written as (see Appendix B)
\BE
\chi\approx \f{4}{\pi} \int_0^{\pi/2} d \theta \bigl((1+\epsilon)\cos \theta
+\epsilon' \sin \theta \bigr)(1+\epsilon)\cos \theta, \label{eq:chi}
\EE
where $\epsilon ' =d \epsilon/d \theta$.
As shown in Appendix B, the order of $\epsilon $ is ${\cal{O}}(L^{-1})$.
Thus the order of the (de)magnification perturbation 
$\chi(L)-1$ is again ${\cal{O}}(L^{-1})$
in the large source size limit $L\gg 1$. 
Note that the analytic formula (\ref{eq:chi}) can be used 
for any type of parity. 

\begin{figure*}
\epsscale{1.0}
\plotone{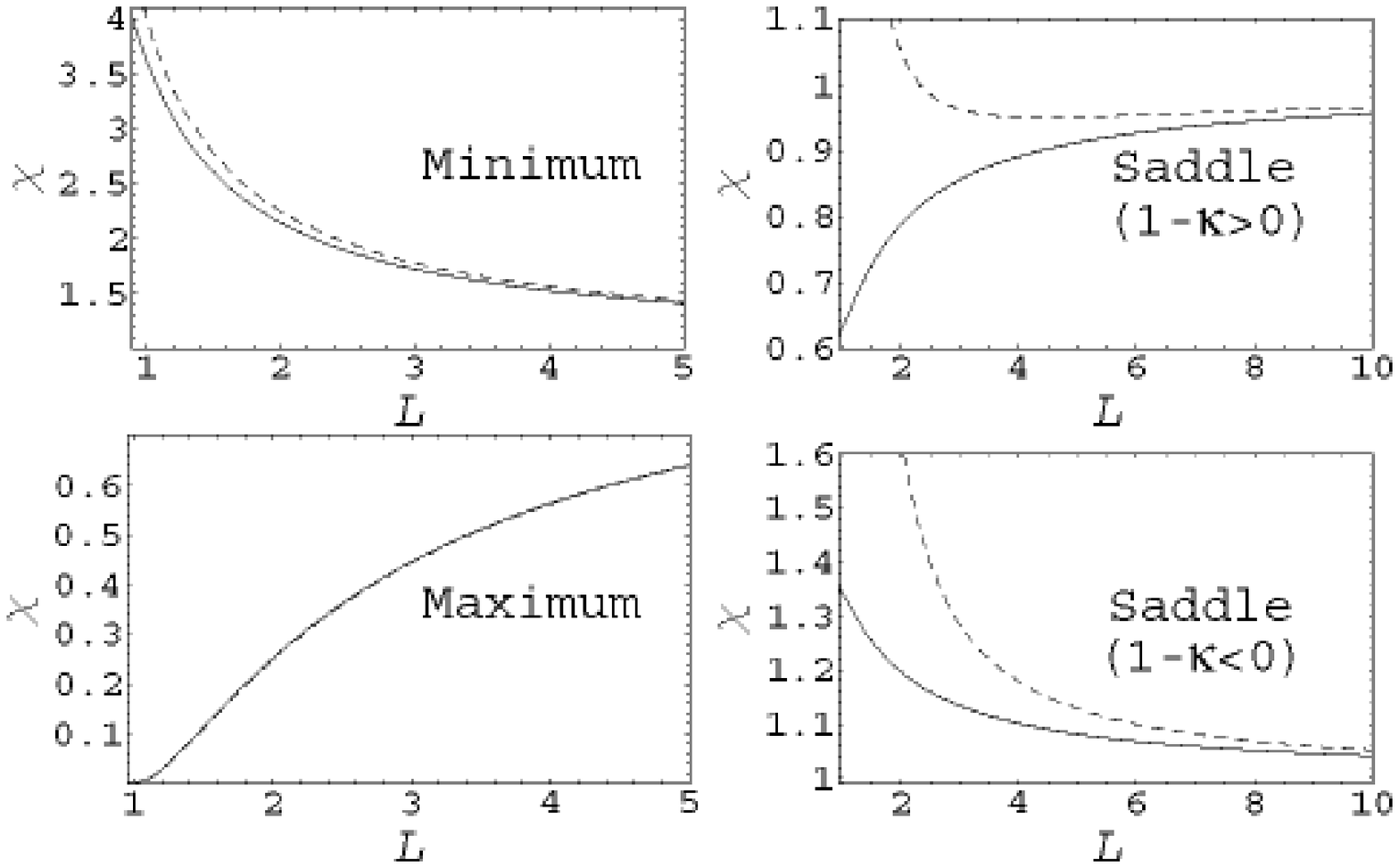}
\caption{ Comparison between numerically calculated 
magnification ratios $\chi$ ({\it{solid curves}}) and 
analytically  calculated ones ({\it{dashed curves}}).
For the maximum ({\it{positive parity}}) and minimum (doubly 
negative parity), 
eq. (\ref{eq:pp-nn}) is used, and for 
saddles (negative parity), eqs. (\ref{eq:epsilon})
and (\ref{eq:chi}) are used.   
Parameters are $(\kappa,\gamma)= $(0.3,0.3) 
({\it {top left}}), (2.5,0.2) ({\it{bottom left}}), 
(0.6,0.6) ({\it{top right}}), and (1.8,2.0) ({\it{bottom right}}). 
The center of an image is placed at $\y=$(0,0) in the source plane. 
The Einstein radius of an SIS lens is set to unity.  }
\end{figure*}
 
As shown in Figure 7, for $L\gtrsim 10$, the analytic formulae (\ref{eq:pp-nn}) and 
(\ref{eq:chi}) for a source centered at an SIS 
give an accuracy to within a few percent 
with respect to the numerically calculated values.  
Because we expanded $\chi$ in $L^{-1}$, the expected 
error at the order ${\cal{O}}(L^{-1})$ is 
$\Delta \chi \sim {\cal{O}}(L^{-2})/(1+{\cal{O}}(L^{-1}))\sim 
(L(L+1))^{-1}$. For instance, we expect an error 
$\Delta \chi \sim 0.2 $ for $L=2$ but 
$\Delta \chi \sim 0.01 $ for $L=10$.  
If the distance between 
the boundaries of the source and caustics is very small, the magnification ratio $\chi$ is sensitive to the position of the lens. 
However, in the large source size limit $L\gg 1$, $\chi$ asymptotically converges to values given by formulae 
(\ref{eq:pp-nn}) and (\ref{eq:chi}). In this case, as 
shown in Figure 8, (de)magnification is less sensitive to the position of the source provided that 
the caustics are totally occulted by the source.
Our analytic formulae are very useful
in representing the effects of 
background parameters $\kappa$ and $\gamma$ on systematic 
(de)magnification in such a case.

As we have seen, for typical
values of convergence $\kappa $ and shear $\gamma$, the 
(de)magnification perturbation $1-\chi(L)$ can be as large as 
$1-\chi(L)\sim 0.4$ at $L\sim 5$ in the positive or 
doubly negative parity case. In other words, even if the 
source size is larger than the Einstein ring or caustics of an SIS
perturber, the (de)magnification effects 
owing to the perturber are still prominent. 
In the negative parity case, the (de)magnification perturbation 
is smaller than for other parities 
$1-\chi(L)\sim 0.1$ at $L\sim 5$ for 
typical values of $\kappa$ and $\gamma$ but it still cannot
be negligible at the order ${\cal{O}}(L^{-1})$.

\begin{figure*}
\epsscale{1.0}
\plotone{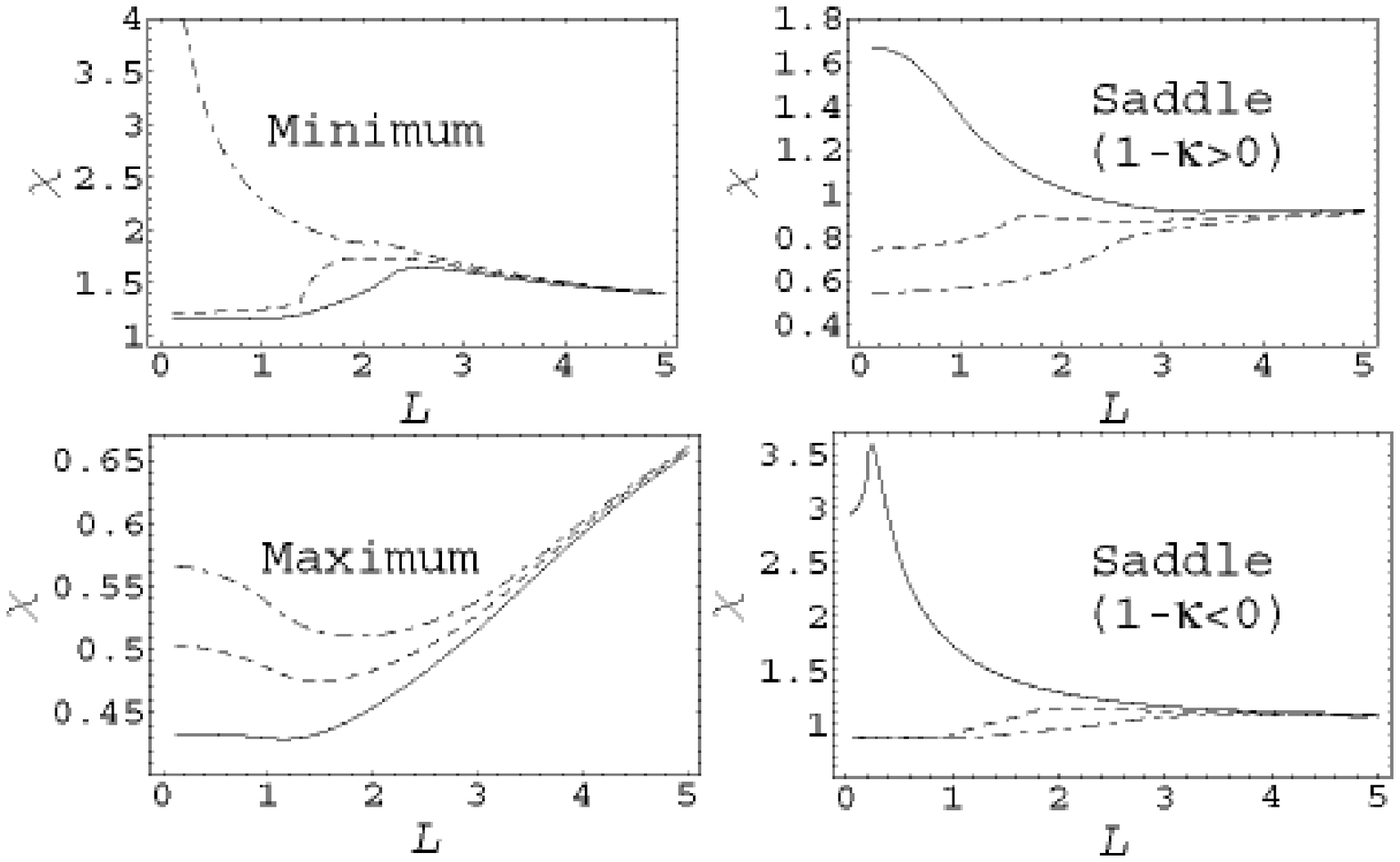}
\caption{Magnification ratios $\chi$
as a function of radius $L$ of a circular top-hat source for   
$(\kappa,\gamma)= $(0.3,0.3) ({\it{top left}}), 
(2.5,0.2) ({\it{bottom left}}),
 (0.7,0.7) ({\it{top right}}), and (1.8,2.0) ({\it{bottom right}}) for 
different image positions 
centered at $\y=$(2,0) ({\it{solid curves}}),  
(1,1) ({\it{dashed curves}}),  (0,2) 
({\it{dash dotted curves}}) in the source plane.  
The Einstein radius of 
an SIS lens is set to unity. An increase
at $L\lesssim 2$ in the minimum and saddle cases 
is owing to a crossing with a caustic.}

\end{figure*}

\subsection{Mean (de)magnification for a point source}
Our results for the systematic (de)magnification effects 
for a circular top-hat source centered at the lens center 
can be applied to the study of the 
mean (de)magnification for a point source
within a radius $L$ from the lens center.

For a point source at $\y=\y_s$ 
in the source plane in which a lens perturber 
is put at the center, the magnification factor corresponding to 
an image $\x=\x(\y_s)$ is written as 
\BE
\mu = \biggl |\f{\del \x}{\del \y }\biggr |_{\y=\y_s}.  
\EE
Then the magnification factor averaged over a disk 
with radius $L$ centered at the center in the source plane
is 
\BE
\langle \mu \rangle_L \equiv \f{1}{4 \pi L^2}\int_{|\y| \le L}d\x.
\EE
This is equivalent to the magnification factor for a
top-hat circular source with radius $L$ centered at the lens center.

The obtained mean magnification factor $\langle \mu \rangle_L $ 
is related to the lensing cross 
section $\sigma(\delta)$ which is defined 
as the cross section
for a magnification perturbation stronger than $\delta$ (Keeton 2003). 
Conversely, we can define the mean magnification 
perturbation for a cross section 
$\sigma=4 \pi L^2$ centered at the perturber as $\langle \delta \rangle_L
\equiv 1-\langle \mu \rangle_L /\mu_{\textbf{bg}}$. The mean
magnification is useful for lensing systems
in which the macro-lensing parameters are precisely determined. 
Comparing an observed value of the perturbation $\delta$ with 
a computed mean perturbation and its variance, one can constrain 
the size of the cross section in units of the area of an Einstein disk
of a perturber, provided that the source size is negligible
in comparison with the Einstein ring.  

Thus, our results are useful not only for estimating
the magnification for extended sources larger than the Einstein ring but
also for estimating the mean magnification 
perturbation for a point source
in substructure lensing (see also Metcalf \& Madau 2001; 
Chiba 2002; Dalal \& Kochanek 2002). 

\section{Measurement of velocity dispersion and distance}
In this section, we study how one can break the degeneracy
between the lensing effects of the velocity dispersion (or total mass)
of a lens perturber and those of its distance along the line of sight
by measuring distortion in the extended images.

Suppose that a halo at redshift $z=z_h$ deflects light from a source
at redshift $z_s$ by an angle $\hat{\al}_h$ and a clump at   
redshift $z=z_c\le z_h$ deflects 
the light by an angle $\hat{\al}_c$ (see 
Figure 9). The normalized lens equations
for an unperturbed system and a perturbed system are
\BE
\y=\x_0-\al_h(\x_0), \label{L1}
\EE
and 
\BE
\y=\x_0-\al_c(\x-\x_c)-\al_h(\x'), \label{L2}
\EE
respectively (Schneider et al. 1992). Here, $\y$ is the position of a 
source, $\x$ is the
position of an image in the microlens plane, $\x_c$ is the
position of a clump, and $\x'$ is the position of 
an intersection between the light ray and the
macrolens plane. The normalized 
angle is defined as $\al_x\equiv
D(z_x,z_s)/D_s \theta_E$, where
$D(z_x,z_s)$ is the angular diameter distance 
between an object at redshift $z_x$ 
and a source at redshift $z_s$, $D_s$ is the
distance to the source, and $\theta_E$ is the Einstein 
radius of the macrolens.

From the lens equation for the microlens system, we have 
\BE
\x'=\x-\beta \al_c (\x-\x_c), \label{L3}
\EE
where the factor 
\BE
\beta=\f{D_{ch}D_s}{D_{cs}D_h} \label{L4}
\EE
encodes a distance ratio written 
in terms of the distance between the clump and halo $D_{ch}$,
the distance between the clump and source $D_{cs}$, and the 
distance to the halo $D_h$. 
 \begin{figure*}[t]
\epsscale{1}
\plotone{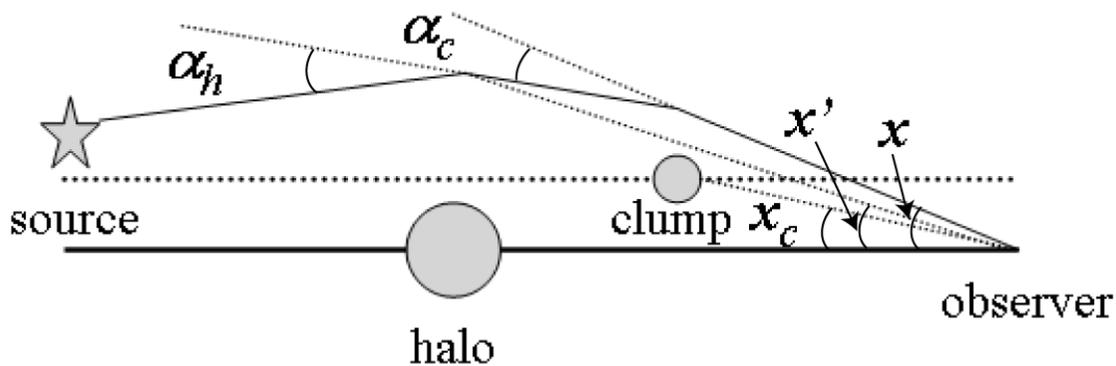}
\caption{Light from a source deflected by a halo
at $z_h$ and a clump at $z_c<z_h$.}
\end{figure*}
In what follows, we
assume that the Einstein radius of the clump is much smaller than that
of the halo, i.e., $|\x'-\x_0|\ll 1$. Then the deflection angle
is approximately given by (Keeton 2003)
\BE
\al_h(\x')\sim \al_h(\x_0)+\bg(\x-\beta \al_c (\x-\x_c)-\x_0), 
~~~\bg=\f{\del \al_h}{\del \x}\biggr |_{\x_0},
\label{L5}
\EE
where $\bg$ represents the convergence and shear for the unperturbed 
macrolens. From equations (\ref{L3}), (\ref{L4}) and (\ref{L5}), we
obtain an effective lens equation for a perturbed system,
\BE
\tilde{\y}=(\bu-\bg_{\textrm{eff}})\tilde{\x}-\al_c(\tilde{\x}),
\EE
where 
\BE
\tilde{\x}\equiv\x-\x_c, \tilde{\y}\equiv-(\bu-\bg)(\bu-\beta \bg)^{-1}(\x_c-\x_0), \label{eq:tildey}
\EE
and
\BE
\bg_{\textrm{eff}}=\bu-(\bu-\bg)(\bu-\beta \bg)^{-1}.
\EE
Here $\tilde{\y}$ is the effective source position that satisfies
$\tilde{\y}=(1-\beta \bg)^{-1}\y$ if the convergence 
$\kappa$ and the shear $\gamma$ 
are constant. 
As a function of the convergence $\kappa$ and shear $\gamma$ 
in the unperturbed macrolens and the distance ratio $\beta$, 
the effective convergence and shear can be expressed as 
\BE
\kappa_{\textrm{eff}}=\f{(1-\beta)
(\kappa-\beta(\kappa^2-\gamma^2))}{(1-\beta \kappa)^2-\beta^2
\gamma^2},\gamma_{\textrm{eff}}=\f{(1-\beta)\gamma}
{(1-\beta \kappa)^2-\beta^2
\gamma^2}. \label{eq:kappa_eff}
\EE
In what follows, we assume that $\bg $ is a constant
matrix that does not depend on the image position $\x$
, i.e., $\al_h=\bg$, and we also assume that an SIS perturber is 
placed at the center of the image plane $\x_c=(0,0)$
, i.e. $\al_c(\x)=\theta_{Ep}~\x/|\x|$, where $\theta_{Ep}$
is the Einstein radius of the SIS perturber.

Now consider an astrometric shift of an image $\Delta \x=\x-\x_0$ 
that is defined as the difference 
between a perturbed macrolensed image position $\x$ and 
an unperturbed macrolensed image position $\x_0$ that satisfies 
$\y=(\bu-\bg)\x_0$.
From equation (\ref{eq:tildey}), one can see that 
an effective image position $\tilde{\x}_0$ that satisfies
$\tilde{\y}=(\bu-\bg_{\textrm{eff}})\tilde{\x}_0$ is equal to the 
unperturbed macrolensed image position $\x_0$. Then the astrometric 
shift of an image with respect to the unperturbed image 
can be written as $\Delta \x=\x-\T{\x}_0$.
Taking coordinates $\x=(x_1,x_2)$ aligned to the shear, astrometric 
shifts of 
images on the horizontal axis $x_1$ are $\Delta \x_{\parallel}
=\theta_{Ep}/(1-\kappa_{\textrm{eff}}-\gamma_{\textrm{eff}})$ and
those of images on the vertical axis $x_2$ are  
$\Delta \x_{\perp}
=\theta_{Ep}/(1-\kappa_{\textrm{eff}}+\gamma_{\textrm{eff}})$. 
Plugging the relations in equation (\ref{eq:kappa_eff})
into these shifts, one can see that the 
distance ratio $\beta$, and the 
Einstein radius $\theta_{Ep}$ of the perturber 
can be determined from observed 
values of astrometric shifts $\Delta \x_{\parallel} $ and 
$\Delta \x{\perp}$ provided that parameters $\kappa$ and $\gamma$ are 
determined from the best-fit macrolens model. 
Furthermore, if the macrolens parameters 
such as $D_s$, $D_h$, and $D_{hs}$ are known, then one can measure the 
distance between the SIS perturber and 
the macrolens from the distance ratio $\beta$ and
the velocity dispersion of the SIS.

The accuracy in the measured distance 
ratio $\beta$ for a given resolution
can be estimated from an astrometric shift $\Delta \x=\x-\x_0$
of a perturbed image with respect to the unperturbed image.
As we have seen, in the coordinates aligned to the background shear, the
$\beta$ dependence of 
the horizontal shift for unperturbed images on the 
horizontal axis $\x_1 $
is $\Delta x_1=\theta_{Ep}(1-\beta (\kappa+\gamma))/(1-\kappa-\gamma) $.
As $\beta$ increases from $0$ to $1$, $\Delta x_1$
decreases by $\Delta (\Delta x_1)=
\theta_{Ep}(\kappa+\gamma)/(1-\kappa-\gamma)$.
Therefore, for macrolenses with 
$(\kappa+\gamma)/(1-\kappa-\gamma)={\cal{O}}(1)$, 
the necessary angular resolution is $\Delta \theta\sim \theta_{Ep} $
in order to determine $\beta$ with a 
relative accuracy $\Delta \beta /\beta<1$.
In other words, $\beta$ can be measured from a 
horizontal shift $\Delta x_1$ with good accuracy
if the angular resolution is comparable to the 
Einstein radius of the SIS perturber.
Similarly, the $\beta$ dependence of 
the vertical shift for unperturbed images on the
vertical axis $x_2$ is $\Delta x_2=
\theta_{Ep}(1-\beta (\kappa-\gamma))/(1-\kappa+\gamma) $.
Therefore, for macrolenses with $\kappa \sim \gamma$,
the $\beta$ dependence of vertical shifts 
$\Delta x_2$ can be observed with an angular 
resolution  $\Delta \theta\sim (\kappa-\gamma)\theta_{Ep} $.
In other words, one needs to resolve an image
that is significantly smaller than the Einstein 
radius of an SIS perturber if $\kappa \sim \gamma$.
However, as we have seen in section 2, 
the Einstein radius $\theta_{Ep}$ can be directly 
measured by the size of the 
dipole structure where the surface brightness 
gradient at the position of the SIS perturber is non 
vanishing (see also section 5). 
Therefore, we conclude that $\beta$ can be measured 
from the observed horizontal astrometric shift $\Delta \x_{\parallel}$
with respect to an unperturbed image if observation with an angular 
resolution $\Delta\theta \sim  \theta_{Ep}$ is achieved. 

In practice, however, we should 
pay attention to various factors that complicate 
the lens system. First, the observed 
distortion can be caused by distortion of the source itself. 
However, this is not a serious problem 
for multiply imaged lensing systems.
By comparing a lensed image with other images, we can easily
distinguish whether the distortion pattern is associated with the  
source properties or not.    
Second, the SIS perturber might not be spherically symmetric.
Third, our assumption of an
SIS density profile for perturbers may not be correct.
Fourth, perturbers may not be single. 
Finally, the center of the perturber may not lie on 
the center of the source.
The second and the third problems can be
solved by measuring the distortion inside the Einstein radius
of the perturber, which is discussed in section 5.
To solve the fourth and the last problems, we 
need a source with spatially
varying surface brightness. Substructures in  
a QSO jet are good examples. If the position of the perturber
is known, we can draw a set of coordinates for which 
the perturber resides at the center. Then we can assign
position vectors to each substructure in the extended source,
which will restrict the lens parameters that can characterize 
the microlens such as $\beta$ and $\theta_{Ep}$.  In other words, 
a sufficient number of substructures of an image 
can restrict the substructure lensing parameters.     
\section{Measurement of mass density profile} 
From a reconstruction of a mapping between a macrolensed 
source image and a microlensed image, we can extract information
of the mass density profile of the perturbers. To do so, we need to resolve distorted images within the Einstein radius
of a perturber. Let us first consider a simple lens model that consists
of a source $S$ with a circular symmetry 
whose surface brightness 
obeys a Gaussian distribution with a standard deviation $\sigma$, 
\BE
f_g(y,\sigma)=\f{1}{\sqrt{2 \pi} \sigma}\exp{\biggl[-\f{y^2}{
2 \sigma^2} \biggr ]}, \label{eq:gauss}
\EE
and a lens (i.e. a perturber) $P$ with a circular symmetry. 
For simplicity, we consider a
polynomially suppressed mass density profile for 
the lens $P$.

First, we consider a system in which a lens is put in the direction
of the line of sight to the center of the source.
Then we have the lens equation
\BE
\y=\x-\f{\x}{|\x|^\alpha},\label{eq:lens-alpha}
\EE
where $\alpha$ corresponds to the power of the polynomially suppressed 
mass density profile of $P$. For instance, 
$\alpha=2$ corresponds to a point mass, and $\alpha=1$ corresponds
to an SIS. From equation (\ref{eq:gauss}) and (\ref{eq:lens-alpha}),
for $\alpha>0$, we obtain the surface brightness profile $f(x)$
in the neighborhood of a lens $x=|x|\ll 1$ as, 
\BE
\ln f(x)\approx -
\f{x^2}{2 \sigma^2 |x|^{2 \alpha}}-\ln{\sqrt{2 \pi}\sigma}. 
\label{eq:lens-approx}
\EE
In Figure 10, one can clearly see the difference in 
the surface brightness of lensed images that depends
on the mass density profile of the lens.
For a lens with $1<\alpha $, the surface brightness 
vanishes toward the
center of a lens as $\ln f(x)\sim -x^{2-2 \alpha}  $. 
This is because the outer dimmed
region of the source is mapped into 
the neighborhood of the center of the lens. 
Consequently, one would observe a ``black hole'' 
at the position of a point mass lens if $\alpha=2$. 
In contrast, the surface brightness
is finite in the neighborhood of the center of an SIS lens. 
In this case,
a region ${R:|y|<1}$ is mapped into
a region ${R':|x|<1}$ on the opposite side (with opposite sign) 
within the Einstein radius but any regions outside the 
Einstein radius are not mapped into 
a region inside the Einstein radius.
Hence, one would see a ``dark hole'' instead of a ``black hole''
for an SIS lens. If the mass density profile of a lens 
is shallower than that of an SIS lens, i.e. $\alpha<1$, then one would see
a bright spot surrounded by a dark hole, because the mass concentration
is so weak that the bending angle for 
a light ray that departed from the brightest region of the source is very small.

In real observations, the detectability of 
these features strongly depends on the resolution
of the map and the linear scale $\sigma$ in which
the surface brightness varies. As $\sigma $ 
gets larger, variability in the surface brightness
of the lensed image within an Einstein radius 
gets smaller as $\propto \sigma^{-1}$ (see Figures 10 
and 11). To determine 
$\sigma$ or equivalently, the
Einstein radius of a perturber 
and $\alpha$, one needs to measure
a mapping between the source plane and the lens plane. 
Suppose that we observe a ``dark hole'' at a peak
in the surface brightness of the source. If there are 
peaks or dips at $y_i >1$ in the source, then 
mapped images of these structures should be observed 
inside dark holes at $x_i<1$.  From more than two
sample points of these structures, one can determine the model parameters
$\sigma$ and $\alpha$.  Even if circular symmetry in the 
projected mass distribution of the perturber is broken, 
a sufficient number of sample points can determine the 
lens parameters such as ellipticity and external shear 
that describe distortion from a circular symmetry.
For instance, in a QSO-galaxy lens system, observation of 
astrometric shifts of subjects in a QSO jet with respect to 
a macrolensed image may determine the model parameters
for perturber lenses. 

Next we consider lens systems in which
the lens is not placed at the peak of the source. 
As shown in Figure 11, 
we would observe a pair of dark and bright spots 
at the position of
the lens where the radial derivative in the surface brightness 
does not vanish. 
For $1 \le \alpha < 2$, the surface brightness
vanishes at the position of the lens because the outer dimmed
region is mapped inside the Einstein radius.
For $\alpha =1$, the surface 
brightness is finite at the position of the lens but
the radial derivative of the surface brightness
at the position of the lens diverges even if the surface brightness 
of the source is smooth.
As in the former case, from more than two sample points, 
one can determine the model parameters $\sigma$ and $\alpha$.   

Our assumption of a Gaussian form of the surface brightness of a source
may be too idealized. In order to estimate the effects of 
deviation from the Gaussian distribution, we consider  
small-scale fluctuations in the 
background source. For simplicity, we added Gaussian fluctuations
with a standard deviation that is 1/15 of the background Gaussian 
peak value and with a vanishing mean to the smooth 
background Gaussian source. 
We calculated lensed images for two examples with the
same smoothed source parameters as in the 
previous cases, namely, $\sigma=2$ and $y=-1$ (large surface brightness
gradient), and $\sigma=10$ 
and $y=-5$ (small surface brightness gradient). 
The minimum wavelength of the Fourier mode of an 
additional fluctuation is assumed to be $\lambda_\textbf{min}=1/4$.
As shown in Figure 12, a discontinuous change in the surface brightness for a point mass  
is still apparent for both cases, whereas for an SIS, 
such a discontinuous change is less apparent for the case with a 
small-scale brightness gradient ({\it{right}}). 
This is because for an SIS,   
an observable astrometric shift is limited to images 
within the Einstein ring of the perturber. 
Therefore, in order to measure the density profile of an SIS, we
need a source whose (smoothed) surface brightness gradient
is sufficiently large.

\begin{figure*}
\epsscale{1.1}
\plottwo{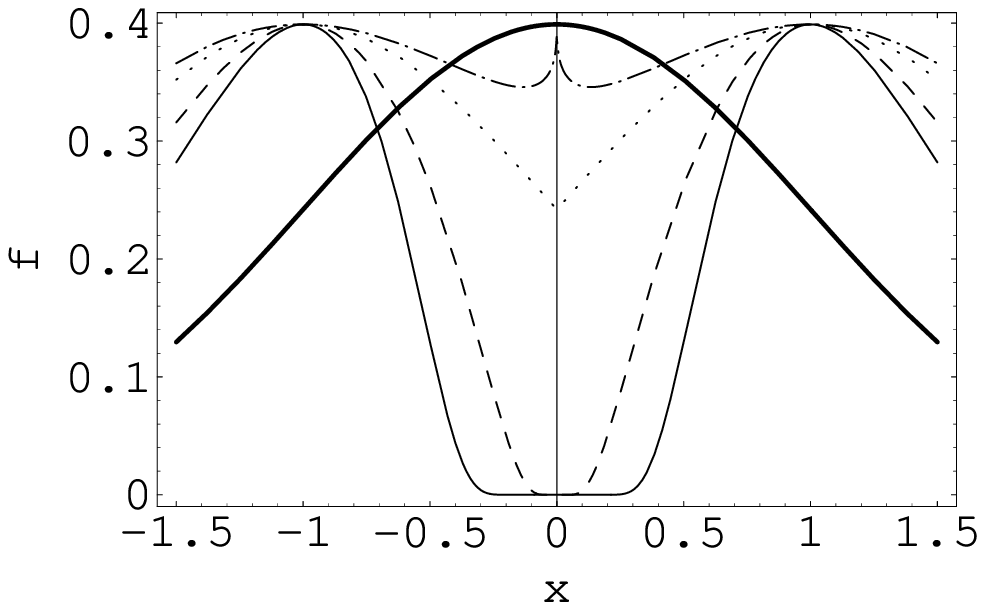}{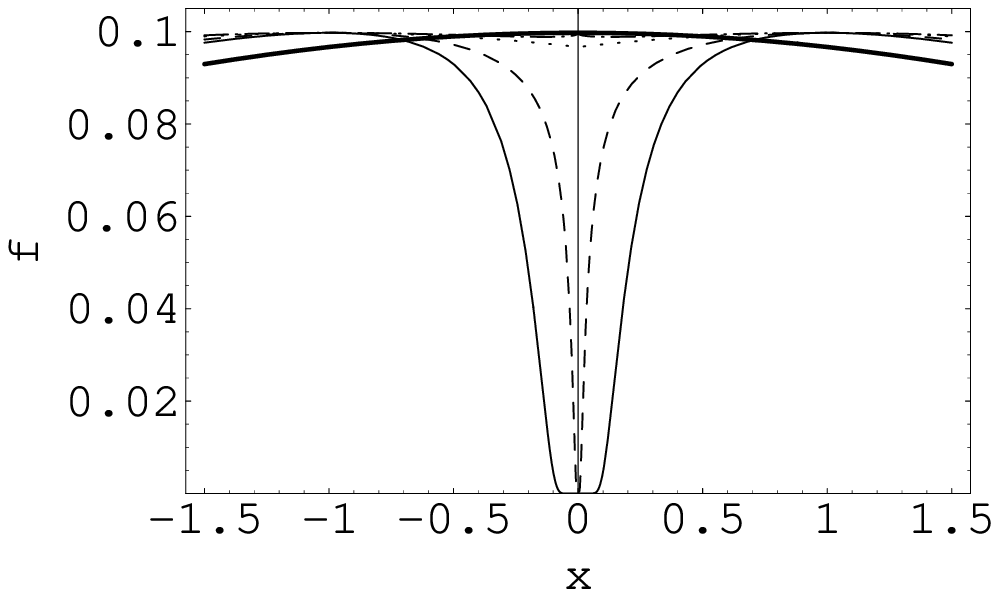}
\caption{{\it{Left}}: Surface brightness of 
lensed images for $\alpha=2$ ({\it {thin solid curve}}), 
1.5 ({\it{dashed curve}}),
1 ({\it{dotted curve}}), and 0.8 ({\it{dash dotted curve}}) 
for a circularly symmetric Gaussian source with
$\sigma=1$ ({\it{thick solid curve}}). The horizontal axis denotes a
coordinate $x$ in the image plane. The Einstein radius of a
 perturber is set to unity. A lens is put at the 
peak of the Gaussian distribution.
{\it{Right}}: Same parameters as in the left panel, 
expect for the standard
 deviation $\sigma=4$. }
\end{figure*}

\begin{figure*}
\epsscale{1.1}
\plottwo{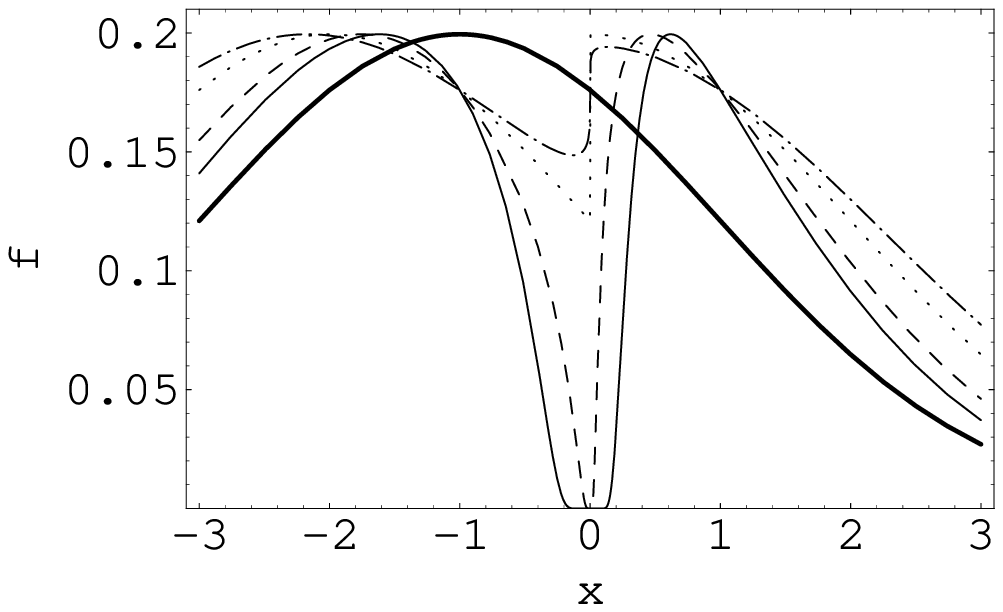}{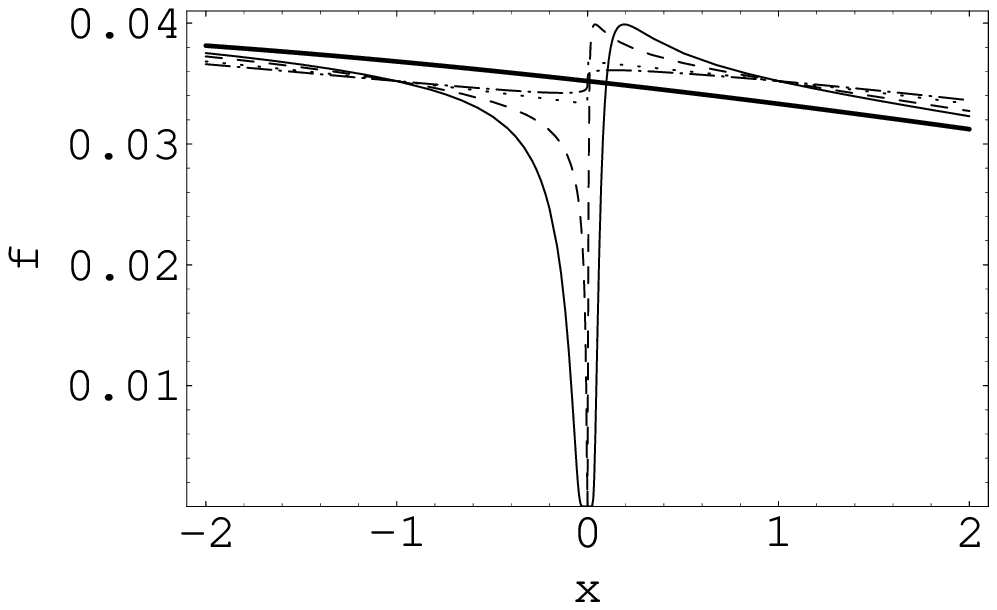}
\caption{Same parameters as in Figure 10 expect for the 
standard deviation  $\sigma $ and the position of the lens. 
{\it{Left}}: $\sigma=2$ and a source is put at $y=-1$ in the source 
plane.
{\it{Right}}: $\sigma=10$ and a source is put at $y=-5$ in the source 
plane.  }

\end{figure*}

\begin{figure*}
\epsscale{1.1}
\plottwo{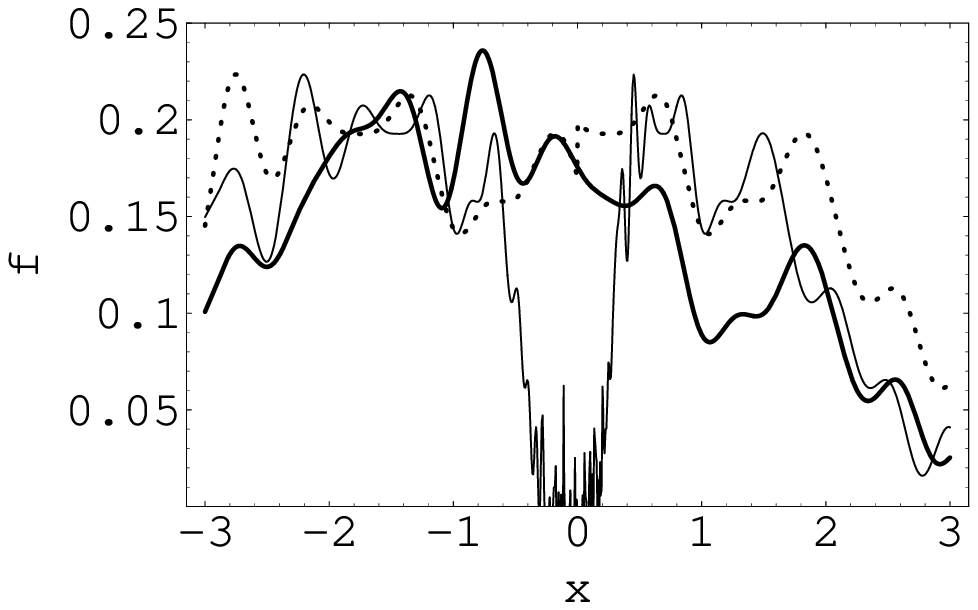}{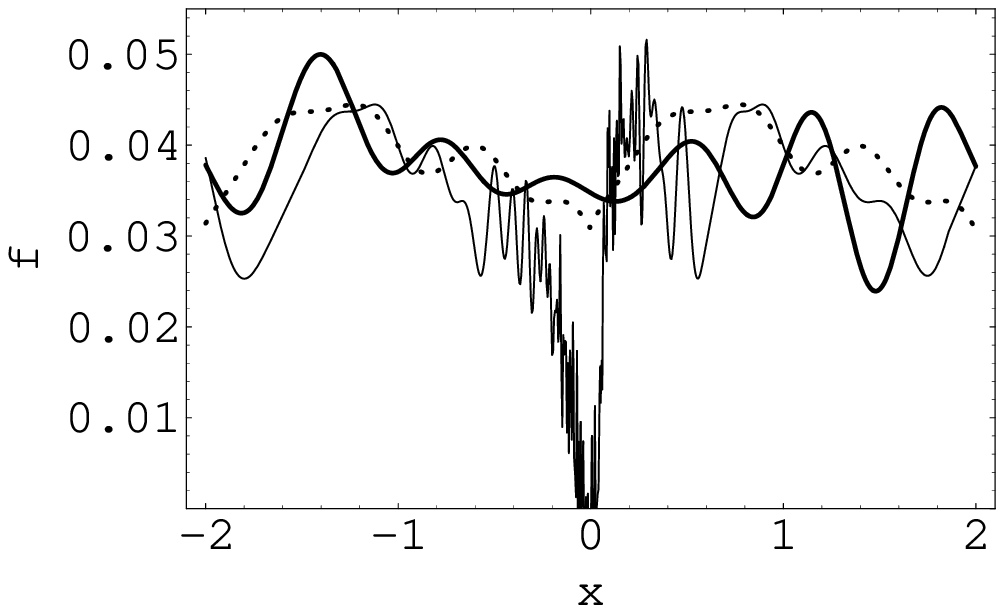}
\caption{Effects of small-scale fluctuations in the surface brightness of
the source.
The parameters are the same as in Fig. 11, except for additional
Gaussian fluctuations with a standard deviation equal to 1/15 of the 
Gaussian peak of the smooth background and a mean equal to 0. 
The minimum wavelength of the
fluctuation is assumed to be $\lambda_\textbf{min}=1/4$.  Thick
solid curves denote the 
surface brightness $f$ of the source, and thin solid curves and dotted
 curves correspond to the surface brightness 
of an image for $\alpha=2$ (point mass) and $\alpha=1$ (SIS), 
respectively.}

\end{figure*}

\section{Summary}
In this paper, we have explored the extended source size effects
in substructure lensing. First, we have shown that 
for a simple SIS lens model with a circular top-hat or Gaussian 
surface brightness profile,  
the magnification effect is prominent even if the 
source size is larger than the size of an Einstein ring.
Second, we have analyzed SIS lens systems with 
background shear $\gamma$ and convergence $\kappa$ 
relevant to substructure
lensing. We have derived analytic (de)magnification asymptotic formulae 
in the large source size limit for three types of 
parity of an image, namely, positive,
negative, and doubly negative parities. 
In the positive (doubly negative) parity case, 
the source is always magnified (demagnified) 
in comparison with the unperturbed macrolens. In the negative
parity case, magnification depends on the sign
of $1-\kappa$. For $1-\kappa>0$, the source tends to be slightly 
magnified, whereas it tends to be slightly demagnified for $1-\kappa<0$. 
Again, we find that in the large source size limit, 
the order of the (de)magnification perturbation $1-\chi(L)$ with respect to
the unperturbed one is typically ${\cal{O}}(L^{-1})$, where $L$ is the 
source radius in units of an Einstein radius.  
We have also shown that these results are relevant to 
the mean (de)magnification for a point source at a distance $L$ from 
the lens perturber center.  
Third, we have shown that one can 
break the degeneracy between the lensing effects of mass and distance of
a substructure based on the distortion pattern in the perturbed 
image provided that the lensing parameters of macrolensing are determined
from the position and flux of multiple images. Finally, we have shown 
that the density profile of a substructure is directly measured
by reconstructing the mapping between a perturbed lens system 
(macrolensing + microlensing) and an unperturbed one (macrolensing only),
which can be achieved by resolving an image
within the Einstein radius of a substructure. 

Although it seems that our results may not be relevant to current
observations of substructure lenses,
they will surely become important tools in future observations
with improved sensitivity and angular resolution. 
For instance, if the sensitivity in the 
measured flux is dramatically 
improved, then we may be able to find an 
asymptotic regime in which the magnification  
perturbation is inversely proportional to the size of the source
with a common center of light.  
Future observations of QSO jets  
in the radio band or of cold dusts in starburst galaxies 
in submillimeter band may reveal such systems if 
observed with different frequencies. If the proper source size
is known, then we can determine the angular size of the Einstein
radius of a lens perturber.       
Similarly, if the angular resolution were dramatically 
improved, then we would be able to get a fruitful amount of information 
from a substructure within a source. For instance, a QSO jet
usually consists of two jet components plus a center nucleus.
Each jet sometimes has some small subjects or small-scale
structures. Then we could determine the spatial variation within the 
source in the global 
linear maps between macro lensed images by comparing their
substructures. Mapping cold dusts around a QSO in a QSO-galaxy
lens at the submillimeter band with an angular resolution of 
$\sim 10$ mas may directly determine 
the distance to a $\sim 10^8-10^9 \ms$ 
subhalo near the center of the lens galaxy (Inoue \& Chiba 2005).
Breaking the mass-distance degeneracy and a 
precise measurement of mass density profile of a perturber 
might both be achieved by such an observation.  
 
In reality, lensing systems are usually much more 
complex and modeled with a large number of 
parameters (often too many!).
For instance, a macrolens can be perturbed 
by a number of subhalos 
with different projected masses, ellipticities, and  
distances from the source.  The corresponding 
best-fit parameters can be determined by 
minimizing a certain measure
between an observed image and a reconstructed image based on 
a certain lensing model. 
 
Many important issues that we have omitted are
(1) irregularity in the source shape, (2) irregularity in
the mass density profile of each perturber, 
(3) time varying source effects, (4) the effect of multiple 
perturbers,  
and (5) lensing parameter error estimates.  
In real observations, we always 
have to tackle all these problems. We will soon address these issues in 
our future work.

%%%%%%%%%%%%%%%%%%%%%%%%%%%%%%%%%%%%%%%%%%%%%%%%%%%%%%%%%
\acknowledgments

We would like to thank S.~Kameno, T.~Minezaki, N.~Kashikawa, and
H.~Kataza for useful discussions on observational aspects of substructure
lensing. 
We would also like to thank the anonymous referee for
useful comments. 
This work has been supported in part by a Grant-in-Aid for
Scientific Research (15540241) of the Ministry of Education, Culture,
Sports, Science and Technology in Japan.

%%%%%%%%%%%%%%%%%%%%%%%%%%%%%%%%%%%%%%%%%%%%%%%%%%%%%%%%%
\appendix
\section{Analytical treatment of overlapping region}
We consider a circularly symmetric top-hat source disk $D$ 
with a radius $L$ located at a distance $\zeta_0$ from the center of an
SIS lens. The magnification factors for an image with 
positive parity and negative parity can be written in terms of 
the radius $L$ and $\zeta_0$ as in equation (\ref{eq:mu3}). 

For $\zeta_0-L>1$, there is no overlapping region
between a source disk $D$ and a unit Einstein ring $E$, 
so the total magnification is $\mu\equiv \mu_++\mu_-=\mu_+$.
For $\zeta_0-L<-1$, an Einstein ring
$E$ is totally included in a source disk $D$, and so we have
$\mu=\mu_++\pi\mu_{-} (L=1,\zeta_0=0)/\pi L^2=\mu_++1/L^2$. 
For $\zeta_0<1$, if a source disk 
$D$ is totally included in an Einstein ring $E$, 
i.e., $L<|1-\zeta_0|$, then $\mu=\mu_++\mu_-$.
These results are summarized in table 1.
%%% Table 1 %%%
\begin{deluxetable}{cccc}
\label{tab1}
\tablecolumns{4}
\tablewidth{3.5in}
\tablecaption{Three Different Cases for Magnification of a
Circular Top-hat Source\tablenotemark{a}}
\startdata
\tableline\tableline
Case   & no overlapping & $D \subset E$  & $E \subset D$   \\
$\mu$  & $\mu_+$ & $\mu_+ + \mu_-$  & $\mu_+ + 1/L^2$
\enddata
\tablenotetext{a}{Here $D$ and $E$ are a source disk
and a unit Einstein ring, respectively.
The definition of
$\mu_+ $ and $\mu_-$ is given in equation (\ref{eq:mu3}).}
\end{deluxetable}

For other cases in which there are some overlaps
between a source disk $D$ and an Einstein ring $E$, 
namely, $|\zeta_0-1|<L<\zeta_0+1$,  
it is difficult to obtain a simple explicit formula.
However, for the cases $L > 1$ and $\zeta_0\gg 1$ 
overlapping some part of an Einstein ring $E$,
a simple approximated formula can be obtained as follows.
In coordinates $(L,\mu(L))$, we may connect 
two points A, $(\zeta_0-1, \mu_+(\zeta_0-1))$, 
and B, $(\zeta_0+1, \mu_+(\zeta_0+1))$, simply 
by a segment, because the area of the overlapping
region $E \cup D$ is a monotonically increasing function of $L$.
However, the derivatives at A and B are not continuous. Instead, 
we can connect A and B by a smooth function
whose derivatives vanish at A and B, because the derivatives
$d\mu/dL $ at $L=\zeta_0-1$ and at $L=\zeta_0+1$ are very small
for $\zeta_0 \gg 1 $.
A simple choice of such a function is
\BE
\delta \mu(s,L,\zeta_0)=\f{\textrm{Erf}(s (L-\zeta_0))
- \textrm{Erf}(-s (L-\zeta_0))+2}{4 L^2}, \label{eq:appf}
\EE
where $s>0$ controls smoothness of $\delta \mu$ around $L=\zeta_0$.
Then we can approximate the magnification factor $\mu$ as
\BE
\mu(L,\zeta_0)\approx \mu_{+}(L,\zeta_0)+\delta \mu(s,L,\zeta_0),~~~ 
L>1,\zeta_0>1. \label{eq:appmu2}
\EE
We find that $s\sim 3$ leads to an accuracy  
within a few percent for $\zeta_0 >1$ (see Figure 1). 
Note that our approximated formula (\ref{eq:appmu2}) 
fails for $\zeta_0<1$ and $L<1+\zeta_0$ because the 
contribution from $\mu_-$ cannot be negligible in that case.
If $\zeta_0 \ll 1 $, then $\mu$ can be approximated as
\BE
\mu(L,\zeta_0) \approx \mu_+(L,0) + \mu_-(L,0) H(1-L),
\EE
where $H$ is a Heaviside step function. 

\vspace{1cm}
\section{Deformation parameters in the negative parity case }
Assuming $\eta \ll 1$, equation (\ref{eq:lens-approx2}) can be solved
as 
\BEA
\epsilon(\theta)&=&(b(1-\kappa+\gamma))^{-1}
\Biggl\{L-b\bigl((1-\kappa+\gamma)-R^{-1}\bigr)-\Biggl
(\cos^2 \theta \bigl(b(1-\kappa+\gamma)LR
\nonumber
\\
&+& a(2 b \gamma-(1-\kappa-\gamma)LR)\bigr)\Biggr )
/R\bigl(a(-1+\kappa+\gamma)
\cos^2 \theta + b(-1+\kappa-\gamma)\sin^2 \theta \bigr)
\Biggr\}, \label{eq:epsilon}
\EEA
and
\BE
\eta(\theta)=\f{\cos \theta \sin \theta \bigl(b 
\xi_2+a(2 b \gamma
/(LR)-\xi_1)\bigr)}{a \xi_1 \cos^2 \theta + b \xi_2 \sin^2 \theta}.
\label{eq:eta}
\EE
In the negative parity case, the approximated solution of $\epsilon
\sim {\cal{O}}(L^{-1})$ for $L \gg 1$ is 
\BE
\epsilon(\theta) \approx L^{-1} \biggl[-1+2\cos^2 \theta +
\f{\sin^2 \theta}{(1-\kappa+\gamma)F(\kappa,\gamma,\theta)}
+\f{\cos^2 \theta}{(1-\kappa-\gamma)F(\kappa,\gamma,\theta)}
\biggr], \label{eq:e-approx}
\EE
where
\BE
F(\kappa,\gamma,\theta)=\sqrt{\f{\cos^2 \theta}
{(1-\kappa-\gamma)^2} + \f{\sin^2 \theta}{(1-\kappa+\gamma)^2}}.
\EE
Now, we evaluate a mean value of $\epsilon$  
averaged over $\theta$ in the negative parity case. 
For $1-\kappa>0$, we have $|1-\kappa+\gamma|
^{-1}<|1-\kappa-\gamma|^{-1}$. Therefore, the contribution of the 
integrand 
$\cos^2 \theta/(1-\kappa-\gamma)F$ is dominant over 
$\sin^2 \theta/(1-\kappa+\gamma)F$. Note that the amplitudes of
both integrands are 1. Because $\langle -1 + 2
\cos^2 \theta \rangle_\theta=0$, we have $\langle  e \rangle_\theta<0$.  
Thus, the area of the perturbed image tends to become smaller 
than that of an elliptic disk $D$ with a semi-major axis $a$ and
a semi-minor axis $b$ at ${\cal{O}}(L^{-1})$.
For $1-\kappa<0$, we have $|1-\kappa+\gamma|
^{-1}>|1-\kappa-\gamma|^{-1}$. Therefore, the 
contribution of the integrand 
$\sin^2 \theta/(1-\kappa+\gamma)F$ is dominant over 
$\cos^2 \theta/(1-\kappa-\gamma)F$, leading to  
$\langle  e \rangle_\theta>0$. 
In this case, the area of the perturbed image 
tends to become larger than that of $E$ (see Figure 6).
In summary, we expect a magnification ratio $\chi > 1$ for $1-\kappa<0 $
but demagnification, $\chi < 1$ for $1-\kappa>0 $. 

As for $\eta$, equation (\ref{eq:eta}) yields
\BE
\eta(\theta)\approx L^{-1} \sin 2 \theta 
\biggl(   1+\f{\gamma}{((1-\kappa)^2-\gamma^2)F(\kappa,\gamma,\theta)}
\biggr). \label{eq:eta2}
\EE
Because $\gamma/((1-\kappa)^2-\gamma^2)F \lesssim 1$, we have  
$\eta \lesssim L^{-1}$. Thus, an assumption of $\eta \ll 1$ is correct
for $L \gg 1$. 
\vspace{1cm}
\section{Deformation parameters in the positive and
 doubly negative parity cases}

To evaluate the accuracy of equation (\ref{eq:pp-nn}), we calculate
the deformation parameters $\epsilon$ and $\eta$.
In the positive parity case, the semi-major axis and 
semi-minor axis of an ellipse $\T{D}$ 
that represents an approximated shape
of the perturbed lensed image are $a=(L+1)\xi_1^{-1}$
and $b=(L+1)\xi_2^{-1}$. Plugging these values into equation 
(\ref{eq:e-approx}), we obtain
\BE
\epsilon(\theta)\sim L^{-1} 
\Biggl(\f{1-\kappa+\gamma \cos 2 \theta }
{((1-\kappa)^2-\gamma^2) F(\kappa,\gamma,\theta)}-1  \Biggr).
\EE
The order of $\epsilon$ is evaluated in terms of 
the median $(1-\kappa)/((1-\kappa)^2-\gamma^2)>0$ of $F$
as
\BE
\epsilon(\theta)\sim L^{-1} \Biggl(\f{\gamma \cos 2 \theta}
{1-\kappa}  \Biggr).
\EE
Therefore, for $1-\kappa-\gamma \gg 0$, the effect of 
deformation from an ellipse $\T{D}$ is negligible at 
${\cal{O}}(L^{-1})$. 

In the doubly negative parity case, we have 
$a=(L-1)\xi_1^{-1}$ and $b=(L-1)\xi_2^{-1}$.
In a similar manner, one can show that 
\BE
\epsilon(\theta)\sim L^{-1} \Biggl(\f{\gamma \cos 2 \theta}
{-1+\kappa}  \Biggr).
\EE
Therefore, for $-1+\kappa-\gamma \gg 0$, the effect of 
deformation from an ellipse $\T{D}$ is again negligible at 
${\cal{O}}(L^{-1})$.

{}

\end{document}